\documentclass[pra,preprint,showpacs]{revtex4} 

\usepackage{graphicx,bm}

\begin{document}

\title{\bf Time ordering in kicked qubits} 

\author{L. Kaplan$^1$, Kh. Kh. Shakov$^1$, A. Chalastaras$^1$, M. Maggio$^1$,
\\ A. L. Burin$^2$, and J. H. McGuire$^1$}

\affiliation{$^1$ Department of Physics, Tulane University, New Orleans, LA
70118 \\ $^2$ Department of Chemistry, Tulane University, New Orleans, LA
70118}

\date{June 22, 2004}

\begin{abstract}
We examine time ordering effects in strongly, suddenly perturbed two-state
quantum systems (kicked qubits) by comparing results with time ordering to
results without time ordering.  Simple analytic expressions are given for state
occupation amplitudes and probabilities for singly and multiply kicked qubits.
We investigate the limit of no time ordering, which can differ in different
representations.
\end{abstract}


\maketitle

\section{Introduction}

There are two reasons to consider time ordering in kicked qubits.  First, the
behavior of a two-state quantum system interacting with a rapidly changing
external field, i.e. a diabatically changing qubit, may be described
analytically.  While such solutions were examined some forty years ago in the
context of
Landau-Zener transitions in atomic and molecular reactions~\cite{smith},
relatively little attention has been paid to this problem in the context of
more recent work using two-state systems~\cite{rwa} ranging from quantum
computing~\cite{qcomp00,josephson} to quantum control of atomic and molecular
reactions~\cite{qcontrol} to manipulation of matter waves~\cite{zcp04}, where
this class of analytic solutions may be useful.  The second reason is that time
ordering has been used recently~\cite{mc01,gm01,mg03} to formulate an
understanding of time correlation in multi-particle systems (or, in the context
of this paper, systems of interacting qubits).  The central question here is
how one particle (or qubit) is connected with other particles (or qubits) in
the time domain.  This problem has previously been formulated~\cite{gm01} using
second order perturbation theory, where observable time correlations between
different particles arise from time ordering of weak, external interactions in
atomic scattering~\cite{gm03}.  The kicked qubit gives us an opportunity to
study the nature of time ordering for a simple, analytically tractable system
in a strongly non-perturbative regime.

Except for relatively simple $e^{-iE_jt/\hbar}$ phases, where the $E_j$ are
simple eigenvalues of a time independent Hamiltonian, there are only two ways
in which time enters the time evolution of a quantum system.  The first is
through the explicit time dependence of an external interaction $\hat V(t)$,
and the second is through the constraint of time ordering imposed by the time
dependent Schr\"{o}dinger equation itself.  This time ordering, discussed
below, imposes a causal-like constraint that places operators such as $\hat
H(t_n) ..... \hat H(t_2) \hat H(t_1)$ in order of increasing time.  This
confining condition interrelates the influence of the time parameters $t_n
\ldots t_1$.  In second order perturbation theory it has been shown~\cite{gm03}
that the time ordering constraint has negligible effect if either $\hat V(t)$
or its variation with time is sufficiently small over the time of the
experiment (perturbative or constant potential limit) or if the energy levels
of the system before perturbation are all nearly the same (degeneracy limit).
In either case, principal value contributions from energy fluctuations in
short-lived intermediate states vanish~\cite{hnt04}. 

In this paper we formulate the problem of time ordering in a non-perturbative
two-state quantum system, i.e. a qubit.  After describing the basic formalism
in Section~\ref{secbasic}, in Section~\ref{sectimeorder} we define the limit
without time ordering and show that time ordering disappears in either the
constant potential limit or in the limit of degeneracy of the two unperturbed
states, where $\hat H(t')$ and $\hat H(t'')$ commute.  We discuss the
relationship between time ordering and the adiabatic approximation.  Analytic
solutions for singly and multiply kicked qubits are presented in
Section~\ref{seckicked}, with and without time ordering.  In the case of a
single kick, we show how time ordering affects the transition probability from
one state to another.  In the subsequently discussed case of a double kick, any
transition is due entirely to time ordering effects.  We discuss corrections
for pulses of finite duration and in Section~\ref{seccalc} provide calculations
illustrating our results.  We present most calculations and some key formulas
in both the Schr\"{o}dinger and intermediate (or interaction) pictures, and
discuss some differences between time ordering effects in the two pictures.

\section{Theory}

\subsection{Basic formulation}
\label{secbasic}

Consider a two-state system, whose states are coupled by a time dependent
external interaction, e.g. a qubit with ``on" and ``off" states.  The time
dependent Hamiltonian for this system may be expressed as
\begin{eqnarray}
\label{genh0}
\hat H(t) &=& \hat H_0 + \hat V(t) \nonumber \\ &=&
\left[ \begin{array}{cc} -\Delta E/2 & 0 \\ 0 & \Delta E/2 \end{array} \right]
+
\left[ \begin{array}{ccc} 0 & V(t) \\ V(t) & 0 \end{array} \right]  \\ \nonumber
 &=& -{\Delta E\over 2} \sigma_z + V(t) \sigma_x \,,
\end{eqnarray}
where $\Delta E= E_2 - E_1$ is the energy difference of the eigenstates of
$\hat H_0$.  Here $\sigma_x$ and $\sigma_z$ are the usual Pauli spin matrices.

Two simplifying, but removable, assumptions have been made in the second line.
First, we assume that all of the time dependence in the interaction operator
$\hat V(t)$ is contained in a single real function of $t$, which is often
justifiable on experimental grounds~\cite{zhao,aarhus,bruch}.  Secondly, in
this paper we assume for convenience that the interaction does not contain a
term proportional to $\hat H_0$.  Obviously, an interaction operator $\hat V$
having the form of a combination of $\sigma_x$ and $\sigma_y$ (i.e. a complex
time dependent field) is equivalent to the above form after rotation of
coordinates.  As we shall discuss later, there are other choices of how to
separate $\hat H$ into $\hat H_0 + \hat V$, and these choices have
consequences.

For a qubit with ``on" and ``off" states
$\left[ \begin{array}{c} 1 \\ 0 \end{array} \right]$ and 
$\left[ \begin{array}{c} 0 \\ 1 \end{array} \right]$,
the probability amplitudes then evolve according to

\begin{equation}
\label{eqmo}
i\hbar {d \over dt}
\left [ \begin{array}{c} a_1 \\ a_2\end{array} \right]
 = \left[
\begin{array}{cc} -\Delta E/2 & V(t) \\ V(t) & \Delta E/2 \end{array} \right ]
\left [ \begin{array}{c} a_1(t) \\ a_2(t)\end{array} \right]
\,.
\end{equation}
The solution to Eq.~(\ref{eqmo}) may be written in terms of the time evolution
matrix $\hat U(t)$ as
\begin{eqnarray}
\label{amps}
\left [ \begin{array}{c} a_1(t) \\ a_2(t) \end{array} \right] 
 = \hat U(t)
\left [ \begin{array}{c} a_1(0) \\ a_2(0) \end{array} \right] 
 =\left [ \begin{array}{cc}
  U_{11}(t)  & U_{12}(t)  \\
  U_{21}(t)  & U_{22}(t)  
\end{array} \right] 
\left [ \begin{array}{c} a_1(0) \\ a_2(0) \end{array} \right] \ ,
\end{eqnarray}
where an experiment is begun at a time $t=0$ and completed at $t = T_f$. 
Since we assume the two-state system is closed, $P_1(t) + P_2(t) =
|a_1(t)|^2+|a_2(t)|^2=1$.

The time evolution operator $\hat U(t)$ may be expressed here as
\begin{eqnarray}
\label{U}
  \hat U(t)  &=& T e^{- \frac{i}{\hbar} \int^t_0 \hat H(t') dt'} 
	= T e^{- \frac{i}{\hbar}  \int^t_0 \left( -{\Delta E
\over 2} \sigma_z + V(t') \sigma_x \right) dt'} 
\\ \nonumber
	&=& T \sum_{n=0}^\infty \frac{(-i/\hbar)^n}{n!}  \int_0^t \hat H(t_n)
	dt_n ... 
	\int_0^t \hat H(t_2) dt_2 \int_0^t \hat H(t_1) dt_1  \ .
\end{eqnarray}
The only non-trivial time dependence in $\hat U(t)$ arises from time dependent
$\hat H(t)$ and time ordering $T$.  The Dyson time ordering operator $T$
specifies that $\hat{H}(t_i) \hat{H}(t_j)$ is properly ordered: $$T \hat H(t_i)
\hat H(t_j) = \hat H(t_i) \hat H(t_j) + \theta(t_j-t_i) \left [ \hat
H(t_j),\hat H(t_i) \right ] \,.$$ Time ordering imposes a connection between
the effects of $\hat H(t_i)$ and $\hat H(t_j)$ and leads to observable,
non-local, time ordering effects~\cite{zhao,aarhus,bruch} when $\left [ \hat
H(t_j), \hat H(t_i) \right ] \ne 0$.

\subsubsection{Pulses}

In this paper we regard $V(t)$ as having the form of a smoothly varying pulse,
or sequence of pulses, each of duration $\tau$ and peaked at $T_k$.  We define
phase angles
\begin{eqnarray}
\label{angles}
\alpha &=& \int_0^{T_f} V(t') dt'/\hbar \nonumber \\
\beta &=& \tau \Delta E/2 \hbar \nonumber \\
\gamma t &=& t\Delta E/2 \hbar \,.
\end{eqnarray}
The angle $\alpha$ is a measure of the strength of the interaction $V(t)$ over
the duration of a given pulse.  In this paper we are mostly interested in the
non-perturbative regime corresponding to $\alpha \geq 1$, so that substantial
changes in the state occupation probabilities, $P_1$ and $P_2$, may occur.  The
angle $\beta$ is a measure of the influence of $\hat H_0$ during the
interaction interval $\tau$.  The angle $\gamma t$ is the phase accumulation of
the propagation due to $\hat H_0$ over a time $t$.  The diabatic (kicked) limit
corresponds to $\beta \ll 1$, the perturbative limit corresponds to $\alpha \ll
1$, and the adiabatic (slow) limit generally corresponds to $\tau \to \infty$.

\subsection{Time ordering}
\label{sectimeorder}

Since time ordering effects can be defined as the difference between a result
with time ordering and the corresponding result in the limit of no time
ordering, it is useful to specify carefully the limit without time ordering.
Removing time ordering corresponds to replacing $T \to 1$ in Eq.~(\ref{U}).
This corresponds to the zeroth order term in an eikonal-like, Magnus expansion
in commutator terms~\cite{magnus}. In the limit of no time ordering, a
multi-particle time evolution operator factorizes into a product of
single-particle evolution operators~\cite{gm01}.

\subsubsection{Limit of no time ordering}

Replacing $T$ with $1$ in Eq.~(\ref{U}), in the Schr\"{o}dinger picture we
have,
\begin{eqnarray}
\label{U0}
 \hat U(t)  &=& T e^{- \frac{i}{\hbar} \int^t_0 \hat H(t') dt'} \to 
	 \sum_{n=0}^\infty \frac{(-i/\hbar)^n}{n!} \left
        [ \int_0^t \hat H(t') dt' \right ]^n
\nonumber \\
	&=&  \sum_{n=0}^\infty \frac{(-i/\hbar)^n}{n!} 
	 \left[ \hat H_0 t + \int_0^t \hat V(t') dt' \right]^n
	=  \sum_{n=0}^\infty \frac{(-i/\hbar)^n}{n!} 
	 \left[ \left(\hat H_0  + \hat{\bar{V}}\right)t \right]^n \nonumber \\
	&=& e^{- i \hat{\bar{H}}t/\hbar} = \hat U^0(t)  \ \ ,
\end{eqnarray}
where
$$\hat{\bar{V}} t = \int_0^t \hat V(t') dt' = \int_0^t  V(t') dt' \sigma_x
	= \alpha \hbar \sigma_x \,,$$
$\hat{\bar H}= \hat H_0+\hat{\bar V}$, and $[\hat H_0,\hat{\bar{V}}]$ terms are
non-zero.  By expanding in powers of $[\hat H(t''), \hat H(t')]$, it is
straightforward to show that to leading order in $\hat V$ and $\hat H_0$ the
time ordering effect is given by
\begin{equation}
\hat U - \hat U^0 \simeq  -\frac{1}{2 \hbar^2} \int_0^t dt'' \int_0^{t''} dt'
\left [\hat H(t''), \hat H(t')\right ] = -\frac{1}{2 \hbar^2} \left[\hat H_0,
\hat V_0 \right] \int_0^t
dt' (t - 2t')
f(t') \,,
\end{equation}
where $\hat V(t') = \hat V_0 f(t')$.  This leading term disappears if the pulse
centroid $T_k=t/2$ and $f(t')$ is symmetric about $T_k$.  Furthermore, $\hat U
- \hat U^0$ vanishes identically in the special cases of $V(t') = 0$,
  $V(t')=\bar V$, or $\Delta E = 0$, as will be discussed below in
Section~\ref{overlap}~\cite{footnote}.

In general there is no simple analytic form for the exact result $\hat U(t)$.
For the result without time ordering, we have
\begin{eqnarray}
\label{u0pulse}
\hat U^0(t) &=& e^{+i \gamma t \sigma_z-i\alpha \sigma_x} \nonumber \\
&=& \left [ \begin{array}{cc}
\cos \xi +i \gamma t {\sin \xi \over \xi} & - i \alpha {\sin \xi \over \xi} \\
- i \alpha {\sin \xi \over \xi} & \cos \xi -i \gamma t {\sin \xi \over \xi}
\end{array} \right ] \,,
\end{eqnarray}
where $\xi =\sqrt{\alpha^2+(\gamma t)^2}$.  Here we have used the well known
identity~\cite{math} $e^{ i \phi \vec{\sigma} \cdot \hat n } = {\bf 1} \cos
\phi + i \vec{\sigma} \cdot \hat n \sin \phi $, following from $(\vec{\sigma}
\cdot \hat n)^n = 1$ (or $\vec{\sigma} \cdot \hat n)$ if $n$ is even (or odd). 

Similarly, in the intermediate, or interaction, picture,
$\hat U_I(t)=e^{i \hat H_0 t} \hat U(t)$, and one has
\begin{eqnarray}
\label{UI0}
 \hat U_I(t)  &=& T e^{-\frac{i}{\hbar} \int^t_0 \hat V_I(t') dt'} 
	\to e^{-\frac{i}{\hbar} \int^t_0 \hat V_I(t') dt'}
	 \nonumber \\
	&=&  \sum_{n=0}^\infty \frac{(-i/\hbar)^n}{n!} 
	 \left [ \int_0^t \hat V_I(t') dt' \right ]^n
	=  \sum_{n=0}^\infty \frac{(-i/\hbar)^n}{n!} 
	 \left [ \hat{\bar{V}}_I t \right]^n = \hat U_I^0(t)  \,,
\end{eqnarray}
where $\hat{V}_I(t') = e^{i \hat H_0 t'/\hbar} \hat V(t') e^{-i \hat H_0
t'/\hbar}$ and $\hat{\bar{V}}_I t = \int_0^t \hat V_I(t') dt'$.  For a Gaussian
pulse of the form discussed in Section~\ref{seccalc},
\begin{equation}
\bar V_I t = \alpha \hbar e^{-\beta^2}\left[
\sigma_x \cos{2 \gamma T_k}+\sigma_y \sin{2 \gamma T_k} \right ]
\end{equation}
and
\begin{equation}
\hat U_I^0(t)= \left [ \begin{array}{cc}
\cos \left (\alpha e^{-\beta^2}\right ) & -i \sin \left(\alpha e^{-\beta^2}
\right ) e^{-2 i \gamma T_k} \\
-i \sin \left(\alpha e^{-\beta^2} \right )
e^{2 i \gamma T_k} & \cos \left (\alpha e^{-\beta^2}\right )
\end{array} \right ]
\label{u0ipulse}
\end{equation}
as long as the measurement time $t$ is after the completion of the pulse, i.e.
$t - T_k \gg \tau$. 

It has been shown previously~\cite{gm03} that to second order in perturbation
theory, $\hat U_I - \hat U_I^0 \sim \left [\hat V_I(t''), \hat V_I(t')
\right]$, somewhat similar to the commutator in the Schr\"{o}dinger picture
above.  From this we immediately see that time ordering effects do not
appear until second order in a perturbative expansion in $\alpha$.  Again,
$\hat U_I-\hat U_I^0 \to 0$ in the special limits $\hat V_I \to 0$, $\hat V_I
\to \hat{\bar V}_I$, or $\Delta E \to 0$, to be discussed immediately below.
However, we will also find in Section~\ref{singpulse2} that $\hat U_I- \hat
U_I^0$ vanishes in the diabatic limit of a single ideal kick, $V(t) \sim
\delta(t - T_k)$, whereas $\hat U-\hat U^0$ is non-zero.  Thus, in principle,
the definition of the limit of no time ordering depends on the picture
(representation) used.  In the discussion we shall relate this difference to
the gauge choice of how one separates $\hat H$ into $\hat H_0 + \hat V(t)$.  As
shown in calculations presented below, this difference can be negligibly small
under some conditions.  

In both pictures, time ordering effects are associated with the fluctuation of
a time dependent interaction about its time averaged value.  

\subsubsection{Relation to other limiting cases}
\label{overlap}

Now we compare the limit without time ordering with the degenerate, weakly
varying potential, and adiabatic limits.  The connection with the diabatic
(kicked) limit appears in Section~\ref{seckicked}, where we discuss the
analytic solution for $\hat U$ in the case of a short pulse.

We noted above that for a general pulse, there is no analytic solution for
$\hat U(t)$.  However, in the limit when the unperturbed states become nearly
degenerate, i.e. $\Delta  E\ll \hbar /T_f$, we obtain
\begin{eqnarray}
\label{additut}
\hat U(t) \to \hat U^{\rm D}(t) & = & e^{-i \int_0^{t} \hat
V(t')dt'/\hbar}= e^{-i \alpha \sigma_x} \nonumber \\ &=&
\left [ \begin{array}{cc}
 \cos \alpha  &  -i \sin \alpha \\
 -i \sin \alpha  & \cos \alpha  
\end{array} \right ] \,.
\end{eqnarray}
This result may be obtained~\cite{sm03} either from the coupled state equations
of Eq.~(\ref{eqmo}), or from Eq.~(\ref{U}) with $\Delta E\to 0$.  In this
degenerate limit the mathematical complexity of the qubit simplifies
significantly, as may be seen by comparing Eq.~(\ref{additut}) with
Eqs.~(\ref{Uadia}) and (\ref{kickedut}).  Most of the complex time connections
have been removed.  We call this degenerate qubit a dit.  In such a dit if the
phase angle $\alpha$ equals $\pi/2$, then an ``on" state is turned off and an
``off" state is turned on with  probability 1.  Then the dit is further reduced
in complexity to a trivial classical bit.

We notice that Eq.~(\ref{additut}) may also be obtained from
Eq.~(\ref{u0pulse}) by taking $\gamma \to 0$.  Thus, time ordering effects
vanish in the degenerate limit.

A second situation in which time ordering effects generically become small is
the case of a constant or weakly varying potential, $\left |\hat V-\hat{\bar
V}\right| \ll \hbar /T_f$.  Then the full evolution matrix approaches the
non-time-ordered expression given by Eq.~(\ref{u0pulse}) or Eq.~(\ref{UI0}).
Clearly the perturbative limit $\alpha \ll 1$ is a special case of this, but
$\hat U - \hat U^0$ vanishes also for a large average external potential $\hat
{\bar V}$, as long as fluctuations around $\hat {\bar V}$ are small.  Such a
situation may sometimes be addressed more transparently by absorbing the
average part of $\hat {V}$ into $\hat H_0$.  In the extreme case $\hat V=0$,
Eq.~(\ref{u0pulse}) reduces to
\begin{eqnarray}
\label{UV0}
 \hat U^0(t) \to e^{- i \hat{\bar{H}}_0 t/\hbar} = e^{-i \gamma t \sigma_z}
 = \left [ \begin{array}{cc}
 e^{i\gamma t }  & 0  \\
 0  & e^{-i\gamma t }  
\end{array} \right ]  \ \ ,
\end{eqnarray}
which of course agrees with the exact evolution matrix $\hat U(t)$.

In summary, time ordering effects disappear either when (i) $\alpha \ll 1$ or
when (ii) $\beta \ll 1$ ($\gamma T_f \ll 1$) in the intermediate
(Schr\"odinger) picture.  The physics becomes especially simple in the overlap
of regimes (i) and (ii).  There, one easily finds that in the Schr\"odinger
picture, for example,
\begin{equation}
\hat U (t)\simeq \hat U^0 (t)\simeq
\left [ \begin{array}{cc}
1 + i \gamma t  & -i \alpha  \\
-i \alpha  & 1 - i \gamma t
\end{array} \right ]  \,.
\end{equation}
It can be shown that corrections and time ordering effects start at $O\left
(\alpha\gamma^2 t^2 \right)$ and $O \left (\alpha^2\gamma t \right )$ for a
symmetric pulse centered at $T_k =t/2$ (see Eq.~(\ref{pulseexpand})).  The
situation is similar in the intermediate picture, except that time ordering
effects vanish identically at leading order in $\alpha$, and begin at $O \left
(\alpha^2\gamma t \right )$ only.

Strictly speaking, time ordering effects also vanish if (iii) $\left [\hat H_0
,\hat V \right ]=0$; however this situation is of little practical interest
due to the fact that no transition or population transfer is possible.

In addition to the degenerate, perturbative, and diabatic regimes, a fourth
limit exists in which analytic solutions for $\hat U(t)$ are generally
available.  In the adiabatic limit where the external interaction $V(t)$
changes slowly in time~\cite{shore}, it is useful to define the instantaneous
level splitting $\Omega(t) = \sqrt{(\Delta E)^2+4V^2(t)}$ and the accumulated
phase $\theta=\int_0^t \Omega(t') dt' / 2\hbar $. Then
\begin{eqnarray}
\label{Uadia}
\hat U(t) & \to & \hat U^A(t) \\
   & = & \left[\begin{array}{cc}
\cos \theta \cos \phi_-
+i\sin \theta \cos \phi_+
&
\cos \theta \sin \phi_-
-i\sin \theta \sin \phi_+ \\
-\cos \theta \sin \phi_-
-i \sin \theta \sin \phi_+
&
\cos \theta \cos \phi_-
-i \sin \theta \cos \phi_+
\end{array} \right]\,. \nonumber
\end{eqnarray}
Here $\phi_{\pm}=\left(\phi(t) \pm \phi(0)\right)/2$ where
$\phi(t')=\tan^{-1}\left (2V(t')/\Delta E\right)$.
In the special case when $V(t)=V(0)$, relevant for a pulse,
\begin{equation}
\label{Uadia2}
\hat U^A(t) =
\left[\begin{array}{cc}
\cos \theta(t) + i {\Delta E\over \Omega(t)} \sin \theta(t) &
-2i {V(t) \over \Omega(t)} \sin \theta(t) \\
-2i {V(t) \over \Omega (t)} \sin \theta(t) &
\cos \theta(t) - i {\Delta E\over \Omega(t)} \sin \theta(t)
\end{array} \right]\,.
\end{equation}
We note that this solution is similar mathematically to the rotating wave
approximation (RWA), which is widely used to describe atomic transitions using
external fields tuned to frequencies near the resonant transition frequency
between two states~\cite{shore,ae,me}.  Inserting Eq.~(\ref{Uadia}) into
Eq.~(\ref{eqmo}), one finds that the leading correction is small when $\hbar
\dot V(t') \Delta E \ll \Omega^3(t')$.  When the splitting $\Delta E$ of the
unperturbed qubit is not small, i.e. $\Delta E \ge V$, then this adiabatic
validity condition reduces to the Landau-Zener criterion~\cite{lz}, namely,
$\hbar \dot V(t') \ll (\Delta E)^2$.  The leading correction to
Eq.~(\ref{Uadia}) or Eq.~(\ref{Uadia2}) is then given by non-adiabatic
transitions at the avoided level crossings,  where $V(t')$ and thus the level
splitting $\Omega(t')$ goes through a minimum.  More generally, for a pulse
having a smooth shape, such as the Gaussian pulses discussed in
Section~\ref{seccalc}, the criterion $\hbar \dot V(t') \Delta E \ll
\Omega^3(t')$ reduces to the union of $\alpha \ll \beta^2$ and $\beta \ll
\alpha^2$.

Remarkably, the adiabatic regime overlaps both with the degenerate limit (when
$\beta$, $\gamma T_f \to 0$ at fixed $\alpha$) and with the perturbative limit
(when $\alpha \to 0$ at fixed $\beta$, $\gamma T_f$).  Thus, Eq.~(\ref{Uadia})
reduces either to Eq.~(\ref{additut}) or Eq.~(\ref{UV0}) as $\Delta E \to 0$ or
$V \to 0$, respectively.  In these overlap regions, time ordering effects are
small.  More generally, however, time ordering effects in the purely adiabatic
regime ($\alpha \gg 1$ and $\beta \ge 1$ or $\beta \gg 1$ and $\alpha \ge 1$)
are large.  Time ordering effects are also large when $\alpha$ and $\beta$ are
both of order unity, where no simple analytic solutions exist for the full
evolution matrix $\hat U(t)$.

\subsection{Kicked qubits}
\label{seckicked}

Now we consider time ordering in kicked qubits, i.e. the diabatic limit where
$V(t) \sim \delta(t - T_k)$.  First, we present analytic
expressions~\cite{berman,nmr} for a kicked qubit, i.e. a two state system
subject to an external interaction, $V(t)$, that changes rapidly with respect
to $T_{\Delta E} =\pi/\gamma=2 \pi \hbar /\Delta E$, the period of oscillation
of the free system.  The corrections needed for finite pulses will be briefly
analyzed.  We also discuss the extension to multiple kicks, using the double
kick as an example.  Finally, we consider the influence of time ordering in
these kicked systems.

\subsubsection{Single kick}

Here we consider a two-state system where the interaction $V(t)$ may be
expressed as a sudden ``kick" at $t = T_k$, namely $V(t)=\alpha \hbar
\delta(t-T_k)$.  For such a kick the integration over time is trivial and the
time evolution matrix in Eq.~(\ref{U}) becomes
\begin{eqnarray}
\hat U^K(t) &=& e^{i{\Delta E\over 2}(t-T_k)\sigma_z/\hbar}
e^{-i \int_{T_k - \epsilon}^{T_k + \epsilon} V(t') dt' \sigma_x / \hbar}
e^{i{\Delta  E\over 2}T_k\sigma_z/\hbar} \nonumber
\\
\label{kickedut}
&=& \left [ \begin{array}{cc}
e^{i\gamma(t - T_k)}  & 0  \\
 0  & e^{-i\gamma(t- T_k)} 
\end{array} \right ]
\left [ \begin{array}{cc}
\cos \alpha   & -i \sin \alpha  \\
 -i \sin \alpha  & \cos \alpha 
\end{array} \right ]
\left [ \begin{array}{cc}
e^{i\gamma T_k }  & 0  \\
 0  & e^{-i\gamma T_k} 
\end{array} \right ]
 \nonumber \\
&=& \left [ \begin{array}{cc}
e^{i\gamma t} \cos \alpha & -i e^{i\gamma (t-2T_k)} \sin \alpha \\
-i e^{-i\gamma  (t-2T_k)} \sin \alpha & e^{-i\gamma t} \cos \alpha
\end{array} \right ] 
\end{eqnarray}
for $t > T_k$.  The second line follows from the identity given below
Eq.~(\ref{u0pulse}) above.  As explained below this solution is valid when
$\beta \ll 1$ so that there is little effect from $\hat H_0$ during the short
time when $\hat V$ is active.

From Eqs.~(\ref{amps}) and (\ref{kickedut}) we have for a kicked qubit
initially found in state $1$:
\begin{eqnarray}
\label{Pk1}
   P_1(t) &=& |a_1(t)|^2 = |U_{11}^K(t)|^2
	= \cos^2 \alpha  \nonumber \\
   P_2(t) &=& |a_2(t)|^2 = |U_{12}^K(t)|^2
	= \sin^2 \alpha  \ .
\end{eqnarray}
The corresponding probabilities for a kicked qubit without time ordering are
discussed below in Section~\ref{seckick0}.

\subsubsection{Finite pulse corrections for a single pulse}

When the pulse width is finite the corrections to Eq.~(\ref{kickedut})  are
$O(\beta)$ and corrections to Eq.~(\ref{Pk1}) are $O(\beta^2)$.  These
corrections result from the commutator of the free Hamiltonian $\hat H_0$ with
the interaction $\hat V$ during the time $\tau$ when the pulse is active.  This
is related to the series expansion that arises in the split operator
method~\cite{splitop}.  For example, in the case of a rectangular pulse of
width $\tau$, the exact time evolution is given by
\begin{equation}
\hat U^{\rm rectangular}(t) = \left [ \begin{array}{cc}
e^{i\gamma t-i\beta} \left (\cos \alpha' + i \beta {\sin
\alpha' \over \alpha'} \right )& -i e^{i\gamma (t-2T_k)} \alpha
{\sin \alpha' \over \alpha'} \\
-i e^{-i\gamma (t-2T_k)} \alpha {\sin \alpha' \over \alpha'} 
& e^{-i\gamma t +i\beta} \left (\cos \alpha' - i \beta {\sin
\alpha' \over \alpha'} \right )
\end{array} \right ] \,,
\end{equation}
where $\alpha'=\sqrt{\alpha^2+\beta^2}$.  To leading order in $\beta$, i.e. in
the width of the pulse, the error in the kicked approximation is given by
\begin{equation}
\label{rectcorr}
\delta \hat U (t)= \hat U^{\rm rectangular}(t)-\hat U^K(t)=
i \beta \left( \frac{\sin \alpha}{\alpha} - \cos \alpha \right )
\left [ \begin{array}{cc} e^{i \gamma t} & 0 \\ 0 & - e^{-i \gamma t}
\end{array} \right ] \,.
\end{equation}

For a narrow pulse having a generic symmetric shape, the leading correction to
the kicked approximation will still have the form
\begin{equation}
\delta \hat U(t) =\hat U(t) - \hat U^K(t)  = i \beta g(\alpha)
\left [ \begin{array}{cc} e^{i \gamma t } & 0 \\ 0 & -e^{-i \gamma t }
\end{array} \right ] \,,
\end{equation}
where $g(\alpha)$ is now a function that depends on the shape of the pulse. 
By comparing Eqs.~(\ref{U}) and (\ref{kickedut}) at leading order
in $\Delta E$, after some algebra one obtains 
\begin{equation}
g(\alpha) = {2 \over \tau} \int dt \, \left [ \cos^2 \left (
\int_{T_k}^t V(t') dt' /\hbar \right )-
\cos^2(\alpha/2) \right ] \,.
\end{equation} 
Expanding $\hat U(t)$ of Eq.~(\ref{U}) and $\hat U^K(t)$ of
Eq.~(\ref{kickedut}) simultaneously in $\Delta E$ and $V$, or equivalently in
$\beta$ and $\alpha$, we find
\begin{eqnarray}
\delta \hat U (t)&=& \hat U(t) - \hat U^K(t)  \nonumber \\ &=&
{i \Delta E \over \hbar}
\int dt \, \left [ \left({\alpha \over 2}\right)^2 - \left (
\int_{T_k}^t V(t') dt' /\hbar \right )^2 \right ]
\left [ \begin{array}{cc} e^{i \gamma t } & 0 \\ 0 & -e^{-i \gamma t }
\end{array} \right ]
\label{pulseexpand}
\\ \nonumber 
&+& {i(\Delta E)^2 \over 2} \int V(t')(t'-T_k)^2 \, dt'/ \hbar^3
\left [
\begin{array}{cc}
0 & e^{i\gamma (t-2T_k)} \\ e^{-i\gamma (t-2T_k)}  & 0
\end{array} \right ]\,,
\end{eqnarray}
so the two leading correction terms scale as $\beta \alpha^2$ and $\beta^2
\alpha$.

\subsubsection{Multiple kicks}
\label{secmultkick}

A series of either identical or non-identical pulses can easily be handled by
multiplication of several matrices of the form of Eq.~(\ref{kickedut}).  For
example, one may consider a sequence of two kicks of opposite sign at times
$t=T_1$ and $t=T_2$, namely, $V_{\rm kick \ antikick}(t)= \alpha \hbar\delta(t
- T_1) -\alpha \hbar \delta(t-T_2)$.  Following the procedure given in
Eq.~(\ref{kickedut}) one obtains the time evolution matrix for $t > T_2$,
\begin{eqnarray}
\label{kickakick}
&&\hat U^{\rm kick \ antikick}(t) = e^{i\gamma(t-T_2) \sigma_z} e^{i\alpha
\sigma_x} e^{i \gamma(T_2 - T_1) \sigma_z} e^{-i\alpha \sigma_x} 
e^{i \gamma T_1 \sigma_z} \\ \nonumber 
&& =  \left[\begin{array}{cc}
e^{i \zeta} ( \cos{ \gamma T_s } 
 + i \sin {\gamma T_s} \cos{2 \alpha} )
& e^{i \gamma (t-2 \bar T)}  \sin {\gamma T_s}  \sin{2 \alpha}  \\ 
  -e^{-i \gamma(t-2 \bar T)}  \sin {\gamma T_s}  \sin{2 \alpha}  &
e^{-i \zeta} ( \cos {\gamma T_s} 
	- i \sin {\gamma T_s} \cos{2 \alpha} ) 
\end{array} \right]\, ,
\end{eqnarray}
where $\zeta = \gamma(t - T_s)$, $\bar T = (T_1+T_2)/2$, and $T_s = T_2 - T_1$.
As $\gamma T_s \to 0$, $\hat U^{\rm kick \ antikick}(t)$ reduces to
Eq.~(\ref{UV0}).

For a double kick with $\bar{V}=0$, we have from Eqs.~(\ref{amps}) and
(\ref{kickakick}),
\begin{eqnarray}
\label{Pkak}
   P_1(t) &=& |a_1(t)|^2 = |U^{\rm kick \ antikick}_{11}(t)|^2
	= \cos^2 \gamma T_s + \sin^2 {\gamma T_s} \cos^2 {2 \alpha}
      \nonumber \\
  \ P_2(t) &=& |a_2(t)|^2 = |U^{\rm kick \ antikick}_{12}(t)|^2
	= \sin^2 {\gamma T_s} \sin^2 {2\alpha}  \ .
\end{eqnarray}

The single-kick result of Eq.~(\ref{kickedut}) remains valid for two or more
kicks of combined strength $\alpha_1 + \ldots + \alpha_n = \alpha$ if the total
phase associated with the inter-kick free evolution, $\gamma(T_n-T_1)$, is
small.  A mathematical analysis for multiple kicks separated by arbitrary time
intervals is straightforward, but not included here.  In the case of a periodic
series of pulses with period $T$, the time evolution may be obtained by
diagonalizing the matrix of Eq.~(\ref{kickedut}) and finding the Floquet
eigenstates and eigenphases.  The two Floquet eigenphases are then given by
$e^{\pm i \chi}$, where $\chi = \cos^{-1} [\cos \alpha \cos {\gamma T} ]$. 

\subsubsection{Time ordering for single and multiple kicks or pulses}
\label{seckick0}

We first consider the case of a single kick or pulse.  In the Schr\"odinger
picture, time ordering effects are present even for a single ideal kick,
specifically the time ordering between the interaction and the free evolution
preceding and following the kick.  Thus, in the absence of time ordering, the
time evolution $\hat U^0(t)$ is given by Eq.~(\ref{u0pulse}), which differs
from the exact expression $\hat U^K(t)$ of Eq.~(\ref{kickedut}) when $\alpha$
and $\gamma t$ are both non-zero.  The time ordering effect $\hat U^K(t)-\hat
U^0(t)$ vanishes in either the degenerate limit $\gamma t \to 0$ or in the
perturbative limit $\alpha \to 0$.  For small $\alpha$ and $\gamma t$ and
assuming $T_k=t/2$, the time ordering effect at leading order takes the form of
a sum of $O\left (\alpha (\gamma t)^2 \right)$ and $O \left ( \alpha^2 \gamma t
\right)$ terms.  For a Gaussian-shaped pulse, the transition probability in the
Schr\"odinger picture without time ordering is given by the second line of
Eq.~(\ref{Pcomp}).

In the intermediate picture, time evolution without time ordering for an ideal
kick is obtained by substituting $\beta=0$ into Eq.~(\ref{u0ipulse}), and
agrees perfectly with the exact expression of Eq.~(\ref{kickedut}), when the
latter is transformed into the intermediate picture.  Thus, time ordering
effects disappear for a single ideal kick in the intermediate picture, in
contrast with the Schr\"odinger case.  This is easily understood by considering
that in the intermediate picture, time ordering is only between interactions at
different times, $\hat V_I(t')$ and $\hat V_I(t'')$, not between the
interaction $\hat V(t')$ and the free Hamiltonian $\hat H_0(t'')$, as in the
Schr\"odinger case.  For a single ideal kick, all the interaction occurs at one
instant, and no ordering is needed.  Of course, for a finite-width pulse, i.e.
$\beta \ne 0$, time-ordering effects do begin to appear even in the
intermediate picture.  To leading order, $\hat U^K_I(t)-\hat U^0_I(t)=
O(\alpha^2 \beta)$.  We note that the time ordering effect in the intermediate
picture is independent of the measurement time $t$, though it does depend on
the pulse width $\tau$ through the $\beta$ parameter.  For a Gaussian-shaped
pulse, the transition probability in the intermediate picture without time
ordering is given by the third line of Eq.~(\ref{Pcomp}).

We are now ready to examine the time ordering effect for a multi-pulse
sequence, focusing on the pulse-antipulse scenario of
Section~\ref{secmultkick}.  In the limit of no time ordering one has $P_2^0(t)
= 0$ in the Schr\"{o}dinger picture as seen from Eqs.~(\ref{amps}) and
(\ref{UV0}) with $\bar{V} = 0$.  In the intermediate picture, however, the
transition probability is non-zero even without time ordering.  For a Gaussian
pulse-antipulse sequence, one may show that $\hat{\bar{V}}_I t = 2 \alpha \hbar
e^{-\beta^2} \sin{\gamma T_s} [\sigma_x \sin {2 \gamma \bar{T}} - \sigma_y\cos
{2\gamma\bar{T}}]$, where $\bar{T} = (T_1 + T_2)/2$.  Note that for $ t-T_2\gg
\tau$, $\bar{V}_I t$ depends on $T_1$ and $T_2$ but not $t$.  Then, $e^{-i
\hat{\bar{V}}_I t/\hbar} = \cos\left[2 \alpha e^{-\beta^2} \sin {\gamma
T_s/2}\right ] - i \left [\sigma_x \sin{2\gamma\bar{T}}  - \sigma_y \cos{2
\gamma \bar{T}} \right] \cdot \sin \left[2 \alpha e^{-\beta^2} \sin {\gamma
T_s}\right]$.  Consequently,
\begin{eqnarray}
\label{U0Idouble}
\hat U_I^0 (t)= \left [ 
\begin{array}{cc} \cos[2 \alpha e^{- \beta^2} \sin {\gamma T_s}]   
&   e^{-2i \gamma\bar{T}} \sin[2 \alpha e^{- \beta^2} \sin {\gamma T_s}]   \\ 
     -e^{ 2i \gamma \bar{T}} \sin[2 \alpha e^{- \beta^2} \sin { \gamma T_s}] 
& \cos[2 \alpha e^{- \beta^2} \sin {\gamma T_s}] 
\end{array} \right ] \ \ .
\end{eqnarray}
We use this result in the next section to study the effect of time ordering on
the transfer of population from one state to another.

We note that as either $\gamma \to 0$ or $\alpha \to 0$, one has $\hat U_I^0(t)
\to \left [ \begin{array}{cc} 1 & 0 \\ 0 & 1 \end{array} \right ] $ in contrast
to $\hat U^0(t) \to \left [ \begin{array}{cc} e^{i \gamma t} & 0 \\ 0 & e^{-i
\gamma t} \end{array} \right ] $.  For simplicity, we may consider the case
where each pulse is an ideal kick, i.e. $\beta=0$.  Then in the perturbative
regime, expanding in $\alpha$, one finds $U_{I\,11}^0 \approx 1 - 2 \alpha^2
\sin^2{\gamma T_s}$, which can be compared to $e^{-i \gamma t} U_{11}^{\rm kick
\ antikick} \approx 1 - 2 i \alpha^2 e^{-i \gamma T_s} \sin{\gamma T_s} $, so
that except for special values of $\gamma$ (including 0), these matrix elements
differ by ${\cal O}(\alpha^2)$.  Similarly, $U_{I\,12}^0 \approx 2 \alpha
\sin{\gamma T_s} e^{-2i\gamma \bar{T}}$ + ${\cal O}(\alpha^3)$ compared to
$e^{-i \gamma t} U_{12}^{\rm kick \ antikick} \approx  2 \alpha \sin{\gamma
T_s} e^{-2i\gamma \bar{T}}$ + ${\cal O}(\alpha^3)$.  Hence $\hat U_I$ and $\hat
U_I^0$ agree to leading order in $\alpha$.  This should not be surprising, as
time ordering has no effect at leading order in perturbation theory.

\section{Calculations}
\label{seccalc}

As an illustrative specific example we present in this section the results of
numerical calculations for $2s \to 2p$ transitions in atomic hydrogen caused by
a Gaussian pulse of width $\tau$.  The occupation probabilities of the $2s$ and
$2p$ states are evaluated by integrating two-state equations using a standard
fourth order Runge-Kutta method.  This enables us to verify the validity of our
analytic solutions for kicked qubits in the limit $\tau \to 0$ and also to
consider the effects of using finite-width pulses.  In this system, the
unperturbed level splitting is the Lamb shift, $\Delta  E= E_{2p} - E_{2s} =
4.37 \times 10^{-6}\,{\rm eV} $.  The corresponding time scale is the Rabi
time, $T_{\Delta E}=2\pi \hbar/\Delta E= 972 \times 10^{-12}$ s, which gives
the period of oscillation between the states.  

For any practical system, the pulse duration $\tau$ can neither be too large
nor too small.  If $\tau$ is larger than $T_{\Delta E}$, then the pulse will
not be sudden and the kicked approximation will fail.  On the other hand if
$\tau$ is too small, then the interaction will have frequency components that
couple the initial state to other levels.  Specifically, if $\tau$ is less than
$2\pi\hbar/(E_{3p} - E_{2s}) \approx 10^{-15}$~s, then the interaction will
induce transitions into the $3p$ level and the system will not be well
approximated by a two-state system.  Also there is another constraint in our
case.  If the experiment lasts longer than the lifetime of the $2p$ state, $1.6
\times 10^{-9}$~s, then we lose population from our two-state system, i.e.
dissipation cannot be neglected.  Similar calculations can be done in many
other applications, including, for example, Josephson
junctions~\cite{josephson}.
 
In the first part of this section we present results for the target state
occupation probability, $P_2$, as a function of time.  We shall examine how
well the approximations we use are satisfied for a $2s - 2p$ transition caused
by a pulse of finite width.  We shall do this first for a single pulse and then
for a double pulse.  In the second part of this section we examine effects of
time ordering.  Here we shall evaluate $P_2(t)$ both with and without time
ordering for pulses of finite width.  This will be done in both the
Schr\"{o}dinger and intermediate pictures.

\subsection{Pulsed two-state system}

In our numerical calculations we use for convenience an interaction of the
form $V(t) = (\alpha\hbar /\sqrt{\pi} \tau) e^{-(t-T_k)^2/\tau^2}$, i.e. a
Gaussian pulse centered at $T_k$ with width $\tau$.  The evaluation of the
integrated pulse strength $\alpha$ in terms of the dipole matrix element for
the $2s - 2p$ transition is discussed in a previous paper~\cite{sm03}.  When
$\tau$ is small enough for the sudden, kicked approximation to hold, $V(t) \to
\alpha \hbar \delta(t - T_K)$, and the analytic expressions of
Eqs.~(\ref{kickedut}) and (\ref{kickakick}) apply.  Here we shall determine how
the occupation probability $P_2(t)$ depends on the pulse width $\tau$, to find
where the kicked results are approximately valid for finite pulses.  We do this
first for a single pulse chosen so that an ideal kick would transfer the
occupation probability $P_2(t)$ suddenly from zero to one at $t=T_k$.  Then we
consider two equal and opposite pulses occurring at times $T_1$ and $T_2$.  We
study this doubly kicked system as a function of both pulse width $\tau$ and
separation interval $T_s=T_2 - T_1$.

\subsubsection{Single pulse} \label{singpulse}

\begin{figure}[ht]
\label{fig_singletime}
{\scalebox{0.7}{\includegraphics{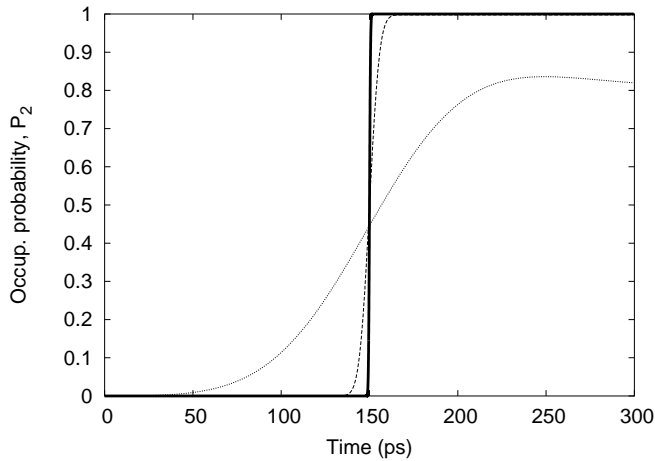}}}
\caption{Occupation probability of the target state as a function of time for a
qubit interacting with a single pulse.  The heavy solid line corresponds to
$\tau=1$ ps (almost an ideal kick), the thin dashed line to $\tau=10$ ps (where
small deviations from an ideal kick occur), and the thin dotted line to
$\tau=100$ ps (where the kicked approximation is breaking down).  The Rabi time
for the oscillation between the states is 972 ps.}
\end{figure}
In Fig.~1 we show results of a calculation for the probability $P_2(t)$ that a
hydrogen atom initially in the $2s$ state makes a transition into the $2p$
state when strongly perturbed by a single Gaussian pulse applied at $t=T_k$.
We have obtained our results by numerically integrating the two-state coupled
equations,
\begin{eqnarray} \label{2s2p-1} 
 i \hbar \dot{a}_1 &=& - \frac{1}{2} \Delta E a_1 
	+ {\alpha \over \sqrt{\pi}\tau} e^{-(t-T_k)^2/\tau^2} a_2 
	\nonumber \\
 i \hbar \dot{a}_2 &=& \ \frac{1}{2} \Delta E a_2 
	+ {\alpha \over \sqrt{\pi}\tau} e^{-(t-T_k)^2/\tau^2} a_1  \ \ .
\end{eqnarray}
Here the pulse is applied at $T_k=150$ ps and we have chosen $\alpha = \pi/2$
so that in the limit of a perfect kick all of the population will be
transferred from the $2s$ to the $2p$ state after $t = T_k$.

In Fig.~1 one sees that the ideal kick results are very nearly achieved by
choosing $\tau$ to be a factor of $10^{-3}$ times smaller than the Rabi time,
$T_{\Delta E}$, in which the population oscillates between the $2s$ and $2p$
states.  When $\tau/T_{\Delta E}\approx 10^{-2}$, a small deviation from an
ideal kick can be seen in the figure.  In this case $P_2(T_f) = 0.9977$.  When
$\tau/T_{\Delta E} \approx 10^{-1}$, the transition takes a few tenths of a
nanosecond to occur and only 82\% of the population is transferred at 300 ps.
The error in $P_2(T_f)$ resulting from the kicked approximation grows as
$(\tau/T_{\Delta E})^2 \sim \beta^2$, as expected from Eq.~(\ref{pulseexpand}).

\subsubsection{A positive followed by a negative pulse}

\begin{figure}[ht]
\label{fig_doubletime1}
{\scalebox{0.7}{\includegraphics{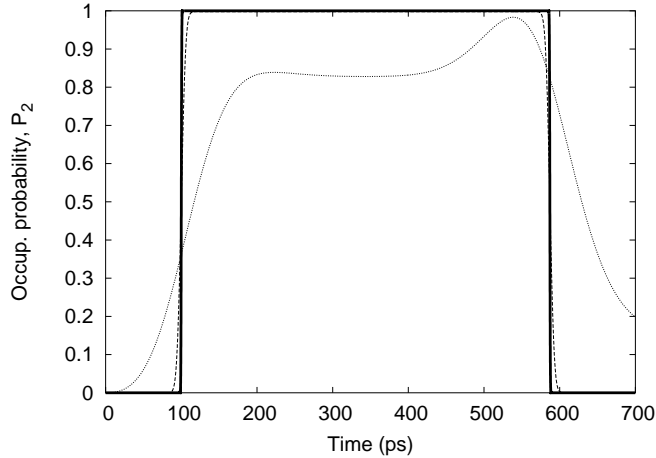}}}
\caption{Occupation probability of the target state as a function of time for a
double pulse that returns the system to its initial state in the kicked limit.
The heavy solid line corresponds to $\tau=1$ ps (almost ideal kicks), the thin
dashed line to $\tau=10$ ps (where small deviations from ideal kicks occur),
and the thin dotted line to $\tau=100$ ps (where the kicked approximation is
breaking down).}
\end{figure}
%
\begin{figure}[ht]
\label{fig_doubletime2}
{\scalebox{0.7}{\includegraphics{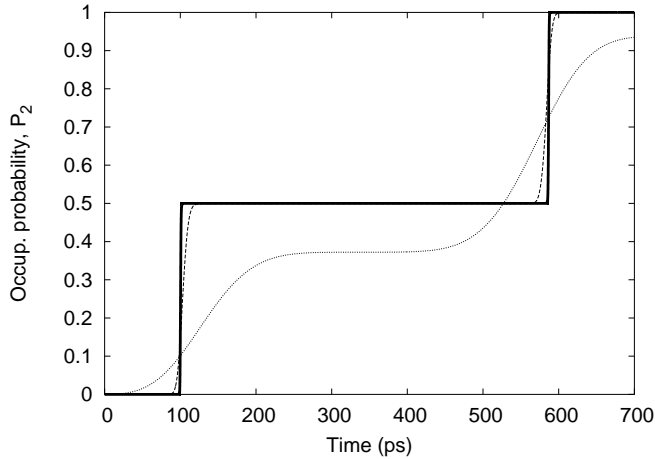}}}
\caption{Occupation probability of the target state as a function of time for a
double pulse that fully transfers population in the kicked limit.  The heavy
solid line corresponds to $\tau=1$ ps (almost ideal kicks), the thin dashed
line to $\tau=10$ ps (where small deviations from ideal kicks occur), and the
thin dotted line to $\tau=100$ ps (where the kicked approximation is breaking
down).
}
\end{figure}

In Figs.~2 and 3 we show results of a calculation for the probability $P_2(t)$
that a hydrogen atom  initially in the $2s$ state makes a transition into the
$2p$ state when acted on by a double Gaussian pulse.  Two Gaussian-form pulses
are applied, the first at $t=T_1$ and the second at $t=T_2$.  The separation
interval between pulses is $T_s=T_2-T_1$.  The final occupation probability of
the target state is measured at $t=T_f$.  The pulses are opposite in sign, but
otherwise identical, so the interaction integrated over the whole interval
$[0,T_f]$ is zero.

Our results for this double pulse have been obtained by numerically integrating
\begin{eqnarray}
\label{2s2p-2}
 i \hbar \dot{a}_1 &=& - \frac{1}{2} \Delta E a_1 
	+ (\alpha/\sqrt{\pi}\tau) \left[e^{-\left (\frac{t-T_1}{\tau}\right)^2}
	- e^{-\left (\frac{t-T_2}{\tau}\right)^2}\right] a_2  \nonumber \\
 i \hbar \dot{a}_2 &=& \ \frac{1}{2} \Delta  E a_2 
	+ (\alpha/\sqrt{\pi}\tau) \left[e^{-\left(\frac{t-T_1}{\tau}\right)^2}
	- e^{-\left(\frac{t-T_2}{\tau}\right)^2}\right] a_1   \ \ .
\end{eqnarray}

In Fig.~2, the first pulse is applied at $T_1=100$ ps and the second at
$T_2=586$ ps, giving a separation time $T_s= 486$~ps~$=\pi/2\gamma$.  From
Eqs.~(\ref{angles}) and (\ref{Pkak}) one sees that this is precisely the value
of $\gamma T_s$ required to yield complete transfer from the $2s$ to the $2p$
state at $T_1$ and then full transfer back to the $2s$ state at time $T_2$ in
the limit of an ideal double kick.  Moreover, we have chosen an integrated
pulse strength $\alpha=\pi/2$ for each pulse for the same reason.  The
parameters for Fig.~3 are the same except that $\alpha = \pi/4$, so that the
$2p$ target state is fully populated after $T_2$ for an ideally kicked system.  

As in Fig.~1 we see that when $\tau/T_{\Delta E}< 10^{-3}$, the kicked limit is
well satisfied, and when $\tau/T_{\Delta E}\approx 10^{-2}$, small deviations
from an ideal kick can be seen in Fig.~2.  In this case $P_2(T_f) = 1.1 \times
10^{-5}$.  Again deviations from the ideal kick limit in the transfer
probability $P_2$ grow as $(\tau/T_{\Delta E})^2$.

The results in Fig.~3 are for a double pulse that first takes the population
halfway from 2s to 2p, and then the rest of the way for an ideal kick-antikick
sequence.  When $\tau/T_{\Delta E}\approx 10^{-2}$, small deviations from an
ideal kick can be observed in the figure near both steps.  At 700 ps,
$P_2(T_f)$ = 0.99934, i.e. the population is nearly, but not quite perfectly
transferred to the target state.  When $\tau/T_{\Delta E}\approx 10^{-1}$, the
transition takes a few tenths of a nanosecond to occur and only 80\% of the
population is transferred, i.e. the transfer is not ideal.  

\subsection{Time ordering}

In this subsection we consider the more complex issue of time ordering in $2s -
2p$ transitions in atomic hydrogen caused by a single or double Gaussian pulse.
Finite-width pulse effects are again considered.  The effect of time ordering
is evaluated by comparing results of calculations with and without time
ordering for the probability $P_2$ of transferring an electron population from
the $2s$ launch state to the $2p$ target state.  Since the limit without time
ordering is different in the Schr\"{o}dinger and intermediate pictures, we
include results for both pictures.

The equations including time ordering are given by Eqs.~(\ref{2s2p-1}) and
(\ref{2s2p-2}) above.  The analogous equations without time ordering are found
by taking $V(t) \to \bar{V}$ in the Schr\"{o}dinger picture and $V_I(t) \to
\bar{V}_I$ in the intermediate picture.

\subsubsection{Single pulse}
\label{singpulse2}

\begin{figure}
\label{fig_single}
{\scalebox{0.5}{\includegraphics{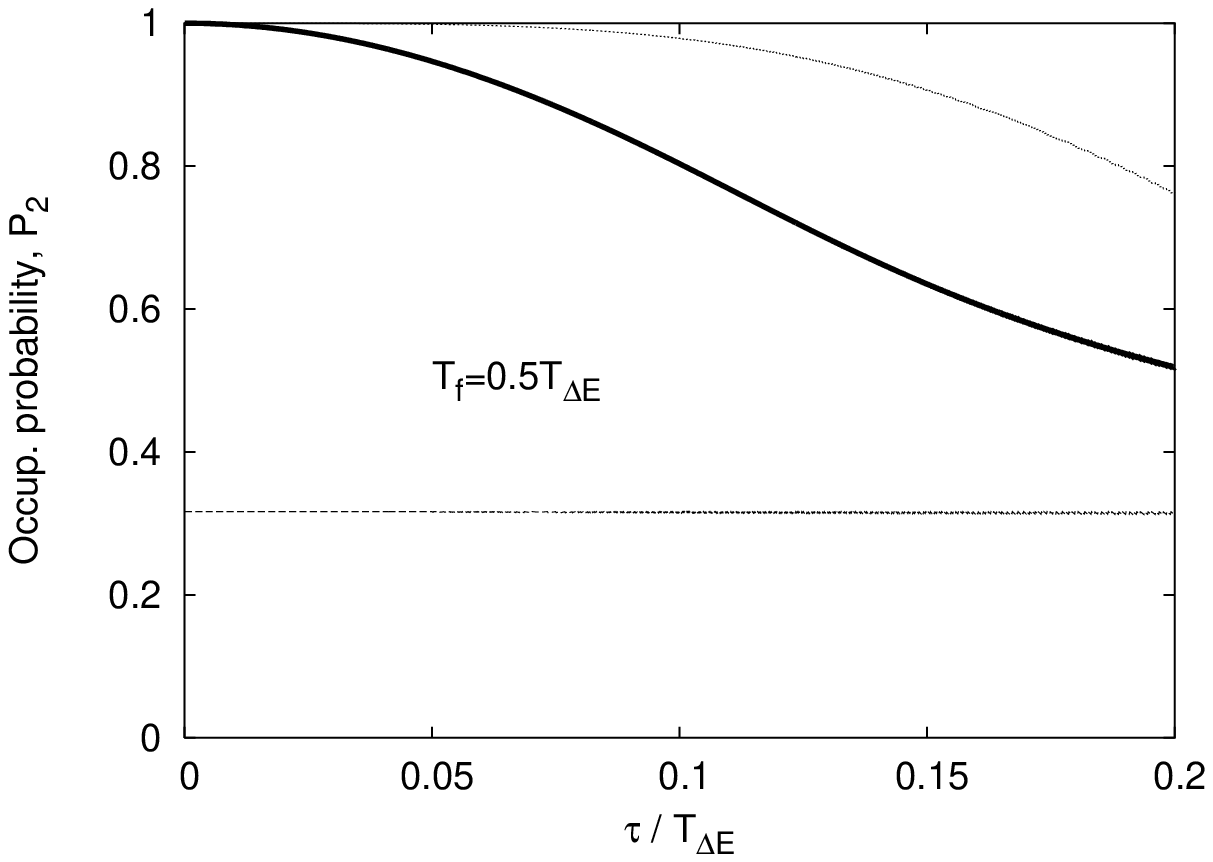}}}
{\scalebox{0.5}{\includegraphics{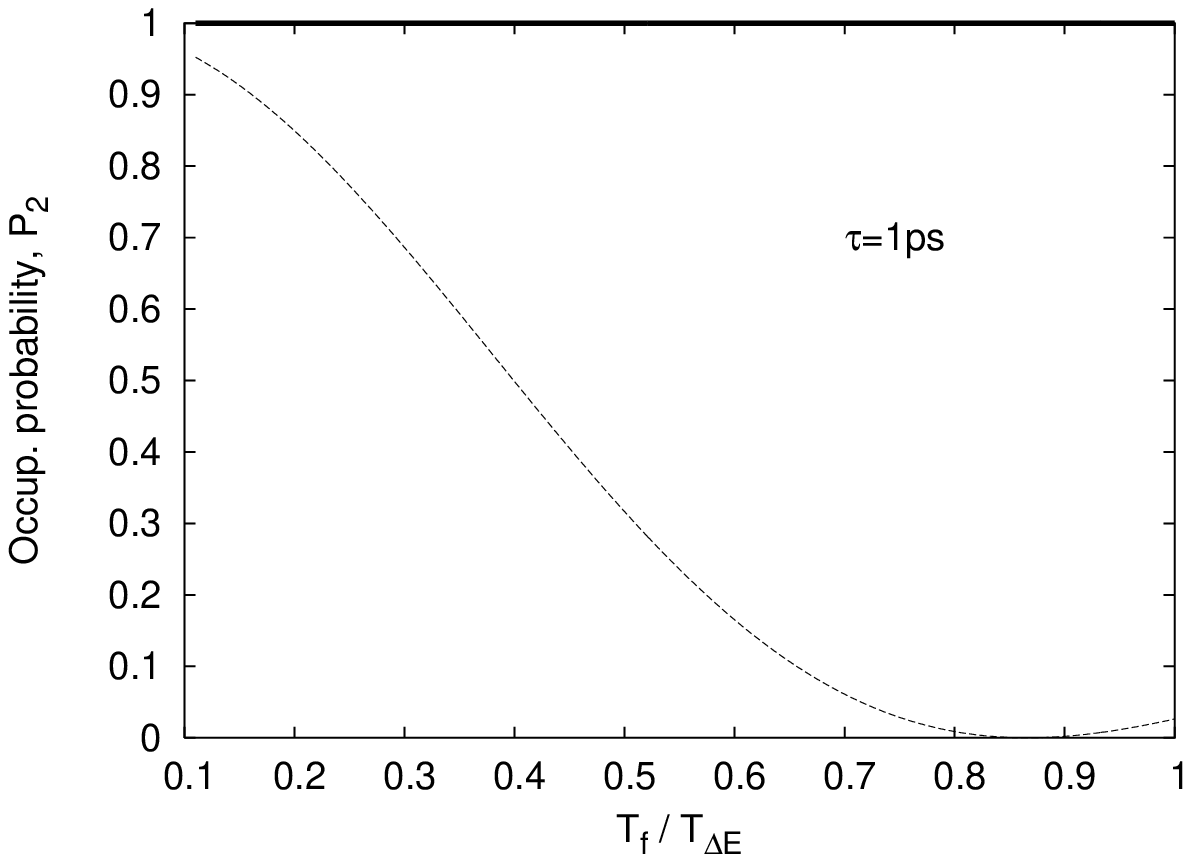}}}

{\scalebox{0.5}{\includegraphics{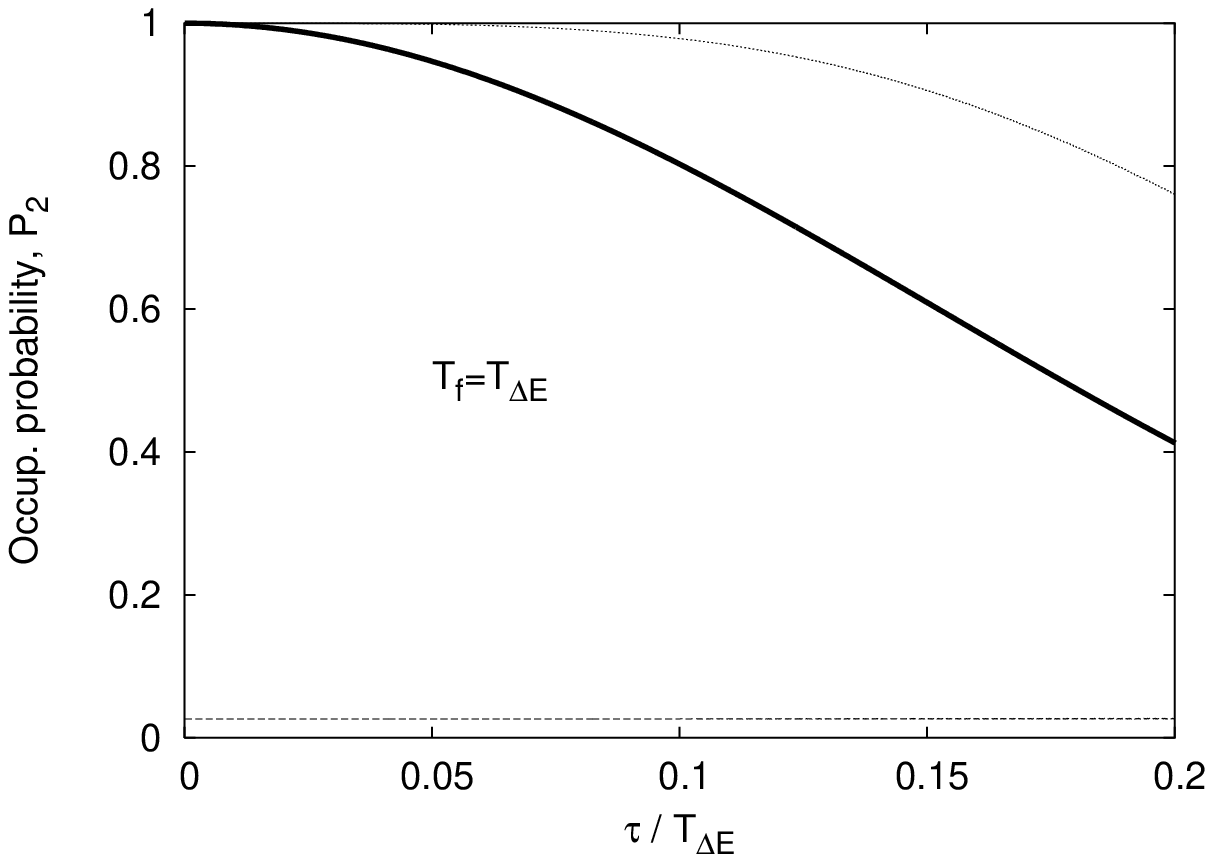}}}
{\scalebox{0.5}{\includegraphics{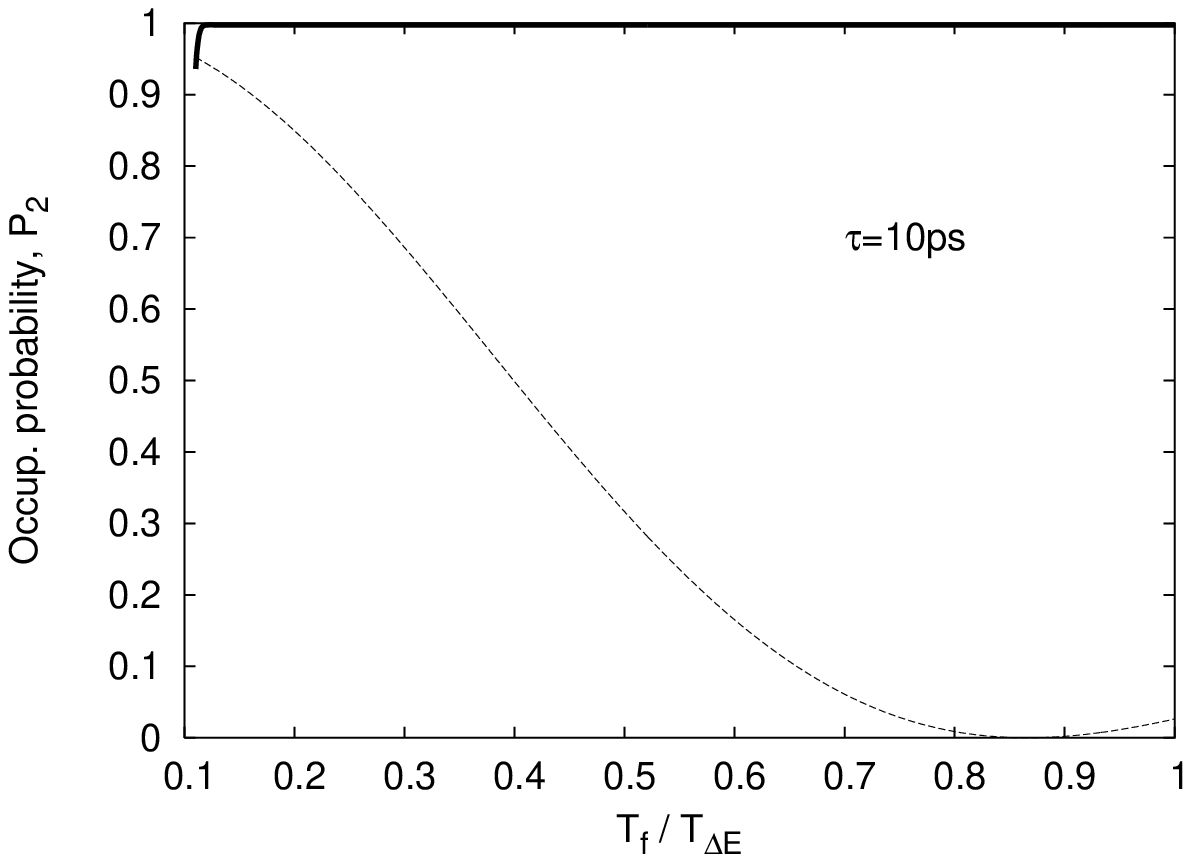}}}

{\scalebox{0.5}{\includegraphics{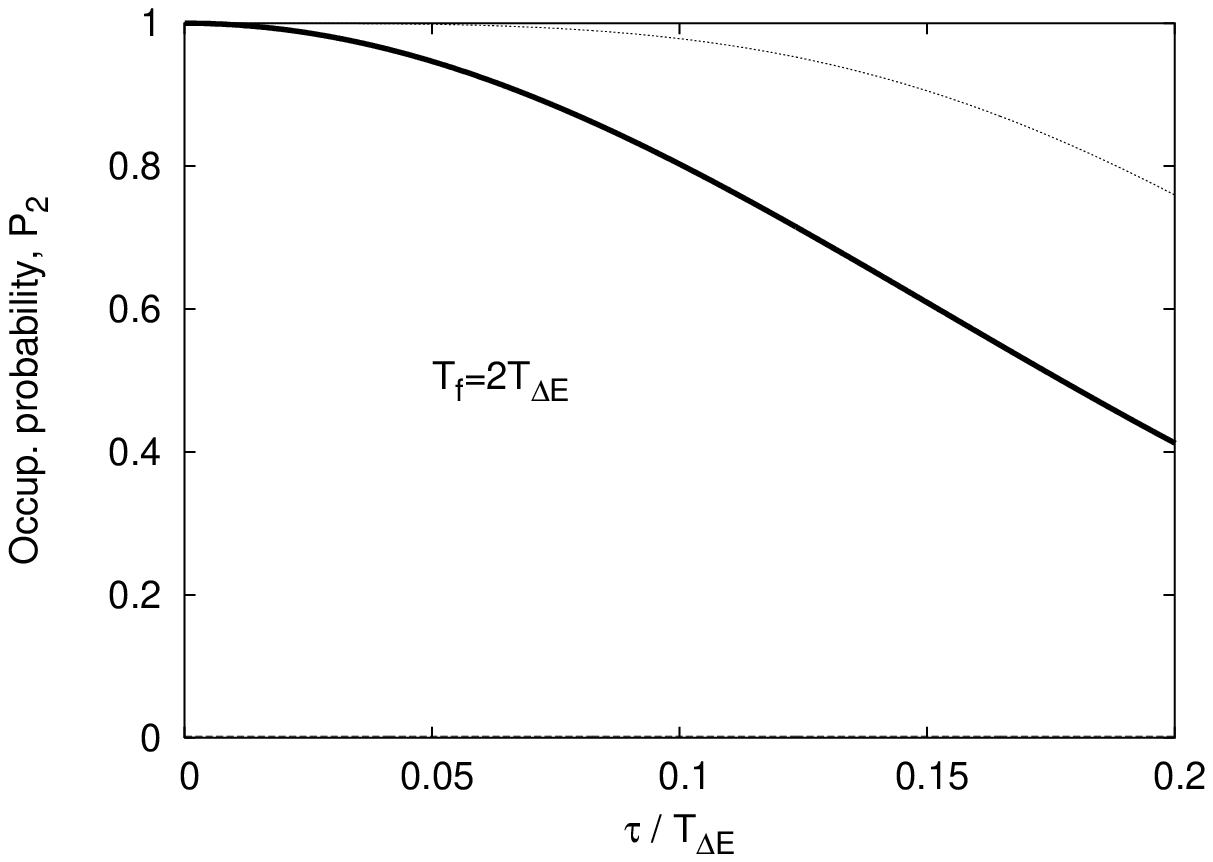}}}
{\scalebox{0.5}{\includegraphics{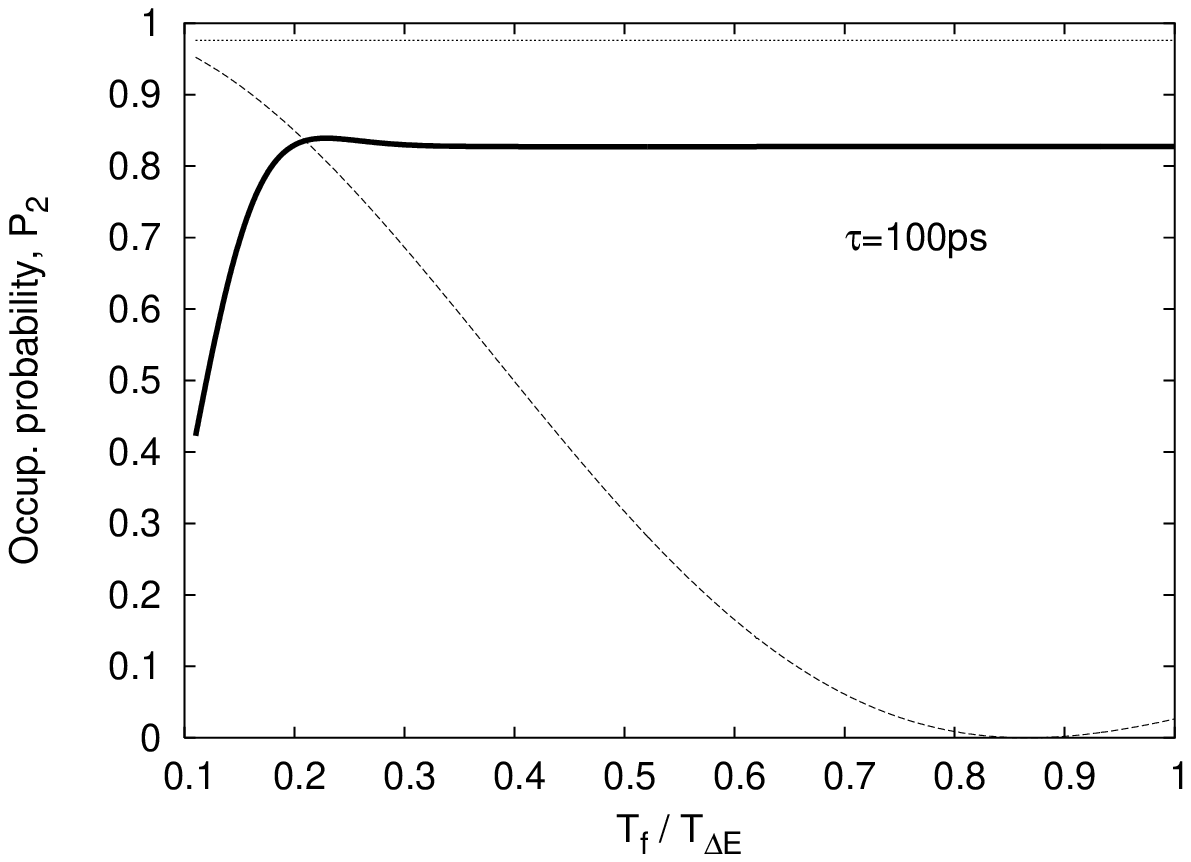}}}
\caption{Target state probability as a function of the pulse width $\tau$ (on
the left), and as a function of the observation time $T_f$ (on the right).
Here $T_{\Delta E}= 2 \pi \hbar / \Delta E=\pi/\gamma$ is the Rabi time for
oscillations between the states, where $\Delta E= E_{2p} - E_{2s}$.  The heavy
line denotes probability including time ordering, the dashed line denotes the
probability in the intermediate picture without time ordering, and the dotted
line represents the probability in the Schr\"{o}dinger picture without time
ordering.  On the right, the lines begin at the midpoint of the pulse,
$T_f=T_k$.  The Schr\"{o}dinger results damp out for large $T_f$ on the right
as explained in the text.}
\end{figure}

In Fig.~4 we show the effects of time ordering for a single pulse.  We have
chosen our parameters so that in the limit of an ideal kick the population is
completely transferred from the $2s$ to the $2p$ state of hydrogen at time
$T_k$, as described above.  On the left hand side of Fig.~4 we show how the
occupation probability of the $2p$ target state varies as a function of the
Gaussian pulse width $\tau$ for three different values of the observation time
$T_f$.  For sharp pulses the exact transfer probability $P_2(T_f)$ and the
transfer probability without time ordering in the intermediate picture $P^0_{I
\, 2}(T_f)$ are quite similar, but differences appear, as expected, when
$\tau/T_{\Delta E}=\beta/\pi$ becomes large. 
 
However in the Schr\"{o}dinger picture there are very large differences between
the results with and without time ordering, $P_2(T_f)$ and $P^0_2(T_f)$, even
for an ideal kick.  This occurs because the energy splitting $\Delta E$ is
non-zero, and for $T_f > \alpha \hbar /\Delta E = \alpha T_{\Delta E}/ 2\pi $,
the average potential $\bar V=\alpha/T_f$ becomes smaller than the energy
splitting $\Delta E$. Thus, for a given pulse, the influence of the potential
necessarily decreases at large $T_f$, and any transfer probability becomes
exponentially small. In effect, the free propagation before and after the pulse
diminishes the effect of the pulse itself in the Schr\"{o}dinger picture, when
time ordering is removed. This behavior contrasts with the intermediate picture
result (Eq.~(\ref{u0ipulse})), where $P_{I \, 2}^0(T_f)$ depends on
$\beta=\gamma \tau$ but not on $T_f$, as seen also on the left side of Fig.~4.
The contrast is evident on the right hand side of Fig.~4 where, after the pulse
has died off, the value of $P_{2}^0(T_f)$ dies out as $T_f$ increases, while
$P_{I \, 2}^0(T_f)$ approaches a constant.

For a single narrow pulse one may compare the time-ordered result for the
transfer probability using the kicked approximation (Eqs.~(\ref{kickedut}) and
(\ref{pulseexpand})) with the exact expressions in the absence of time ordering
in the Schr\"odinger and intermediate pictures, given by Eqs.~(\ref{u0pulse})
and (\ref{u0ipulse}),
\begin{eqnarray}
\label{Pcomp}
	P_2(T_f)   &=& \sin^2 \alpha +O(\alpha^2\beta^2) \nonumber \\
	P_2^0(T_f) &=& \frac{\alpha^2}{\alpha^2 + (\gamma T_f)^2} 
		\sin^2 \sqrt{\alpha^2 + (\gamma T_f)^2} \\ 
	P_{I \, 2}^0(T_f) &=& \sin^2(\alpha e^{-\beta^2}) \nonumber \ \ .
\end{eqnarray}
These three equations are consistent with the numerical results shown in
Fig.~4.  As $\Delta E \to 0$, $\gamma T_f$ and $\beta$ become small, time
ordering effects disappear, and all three results coincide at $\sin^2 \alpha$.
  
\subsubsection{A positive followed by a negative pulse}

\begin{figure}
\label{fig_double}
{\scalebox{0.5}{\includegraphics{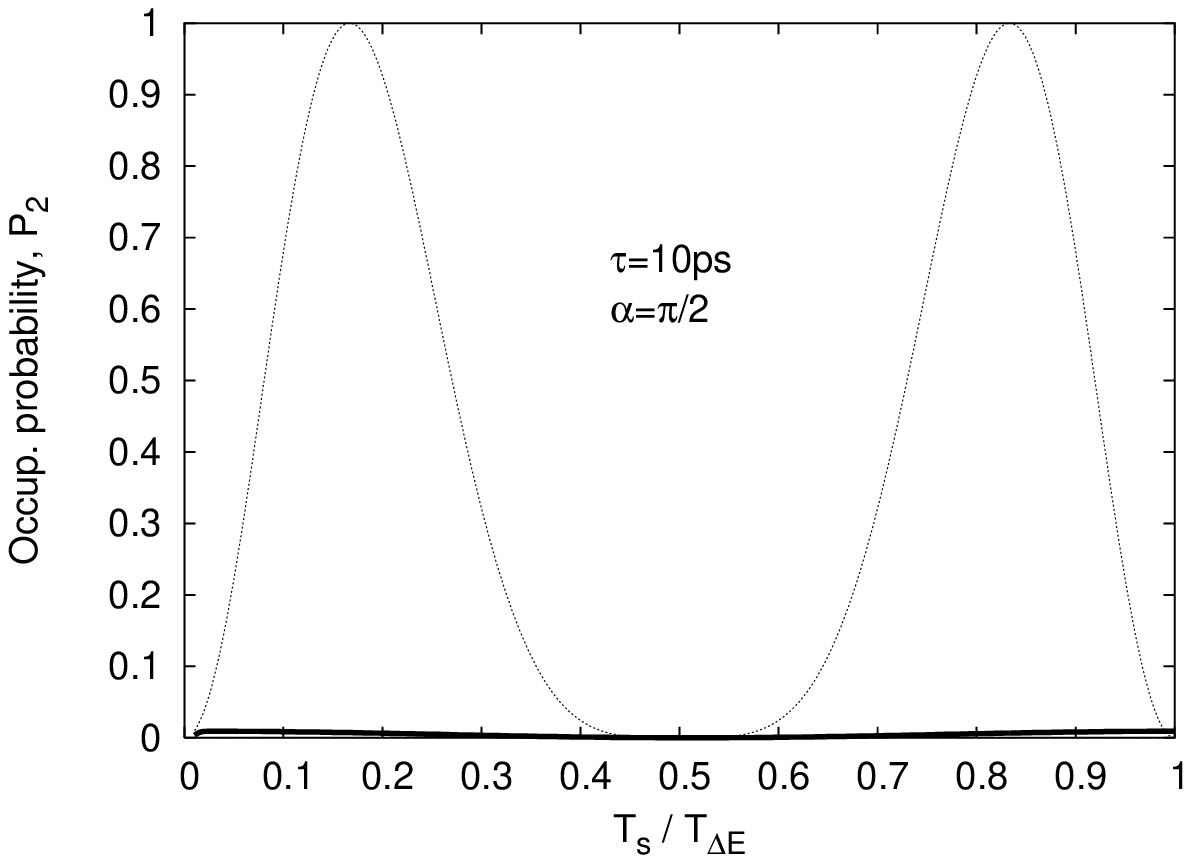}}}
{\scalebox{0.5}{\includegraphics{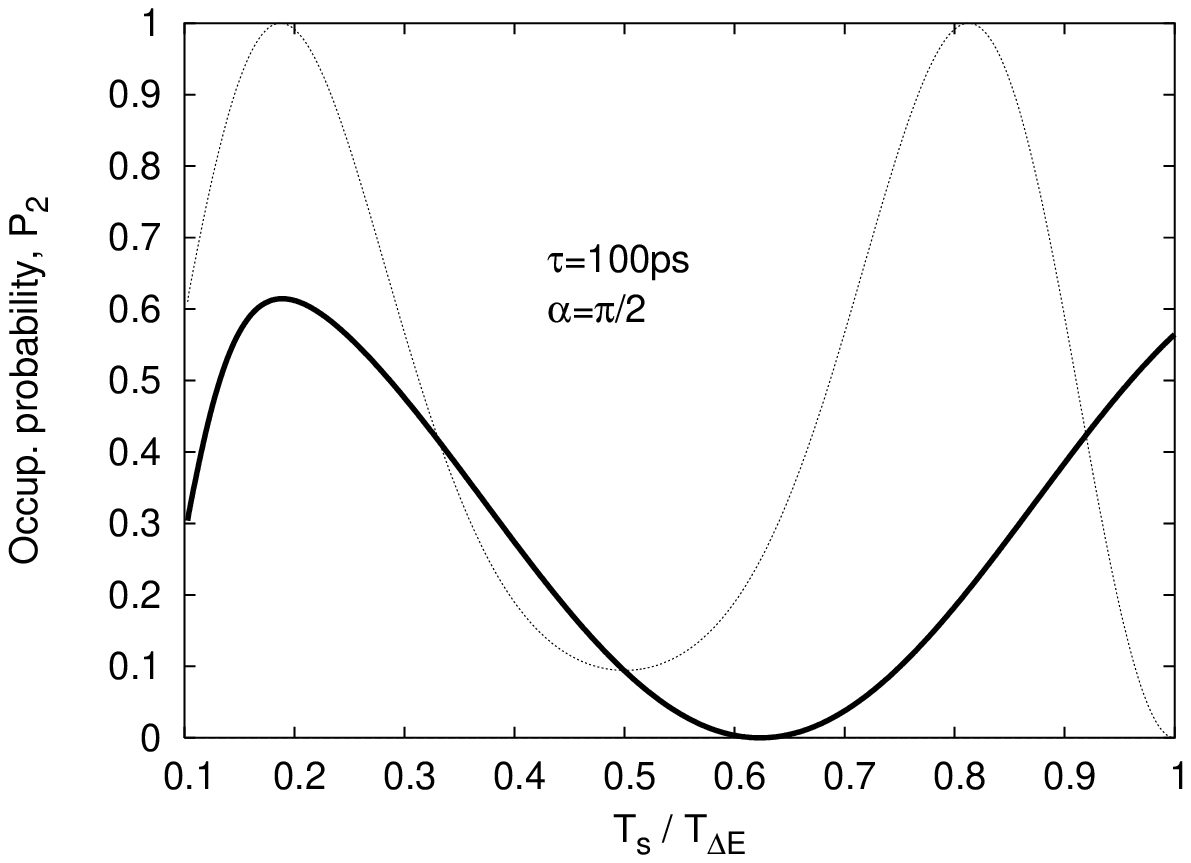}}}

{\scalebox{0.5}{\includegraphics{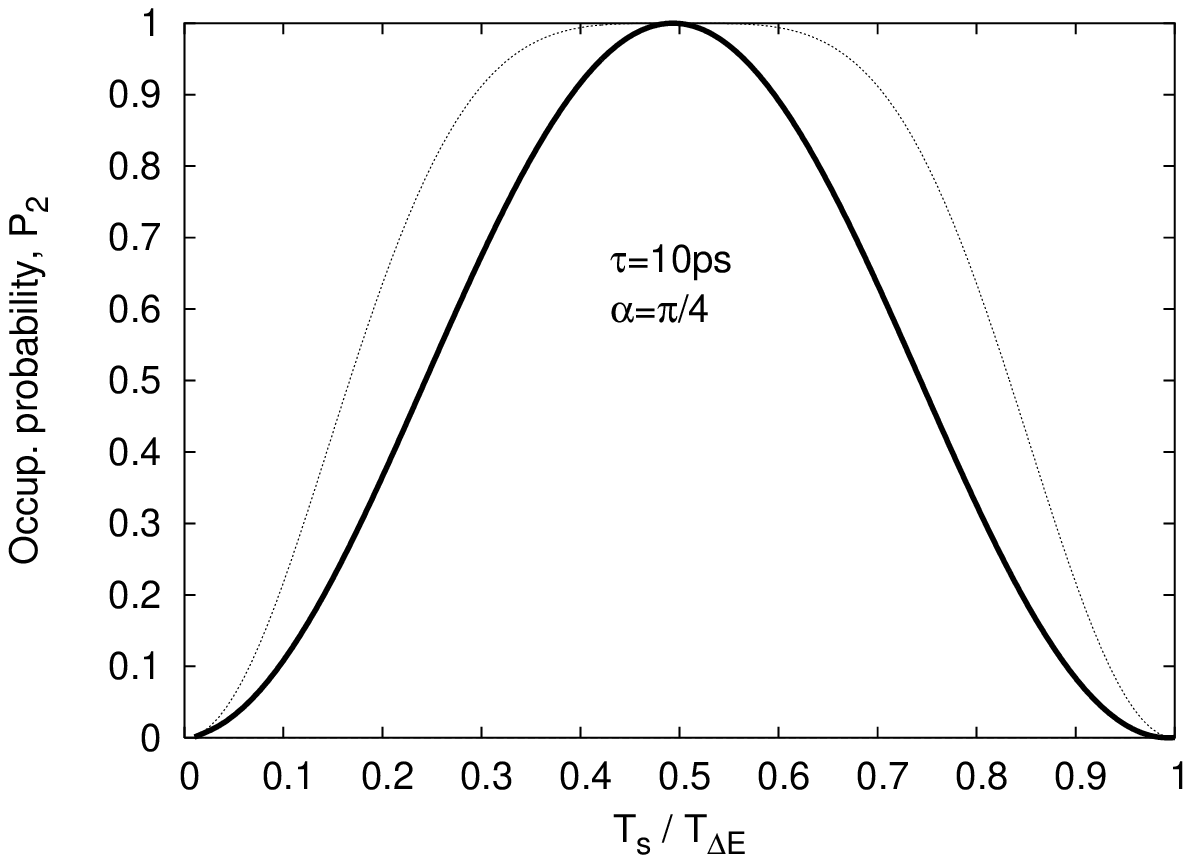}}}
{\scalebox{0.5}{\includegraphics{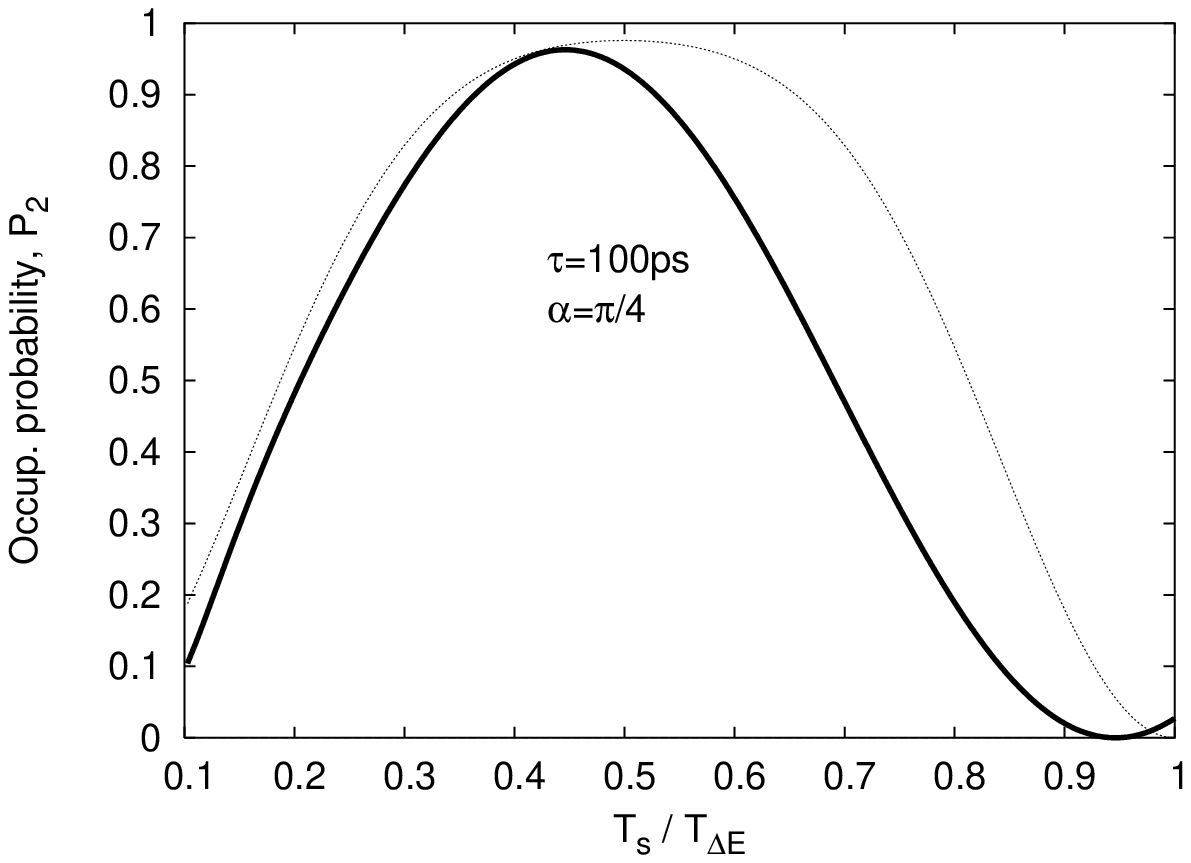}}}

{\scalebox{0.5}{\includegraphics{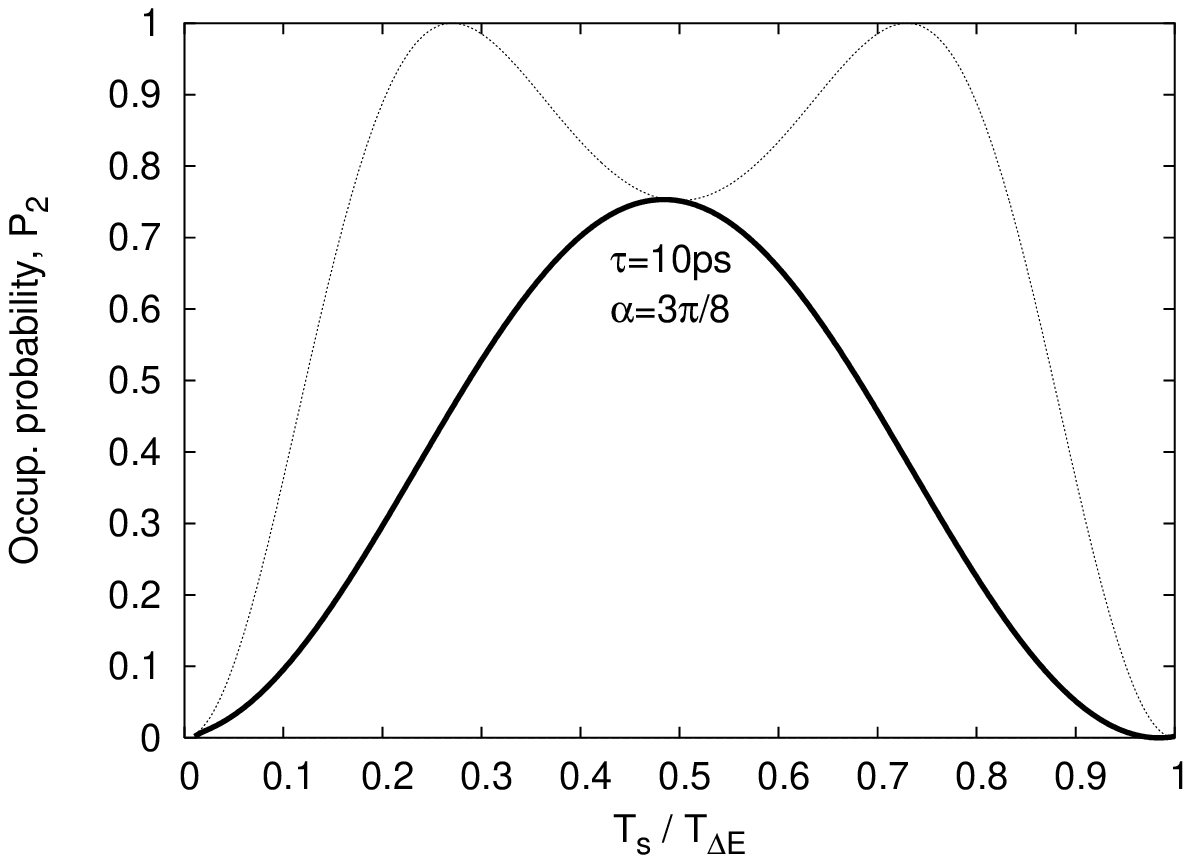}}}
{\scalebox{0.5}{\includegraphics{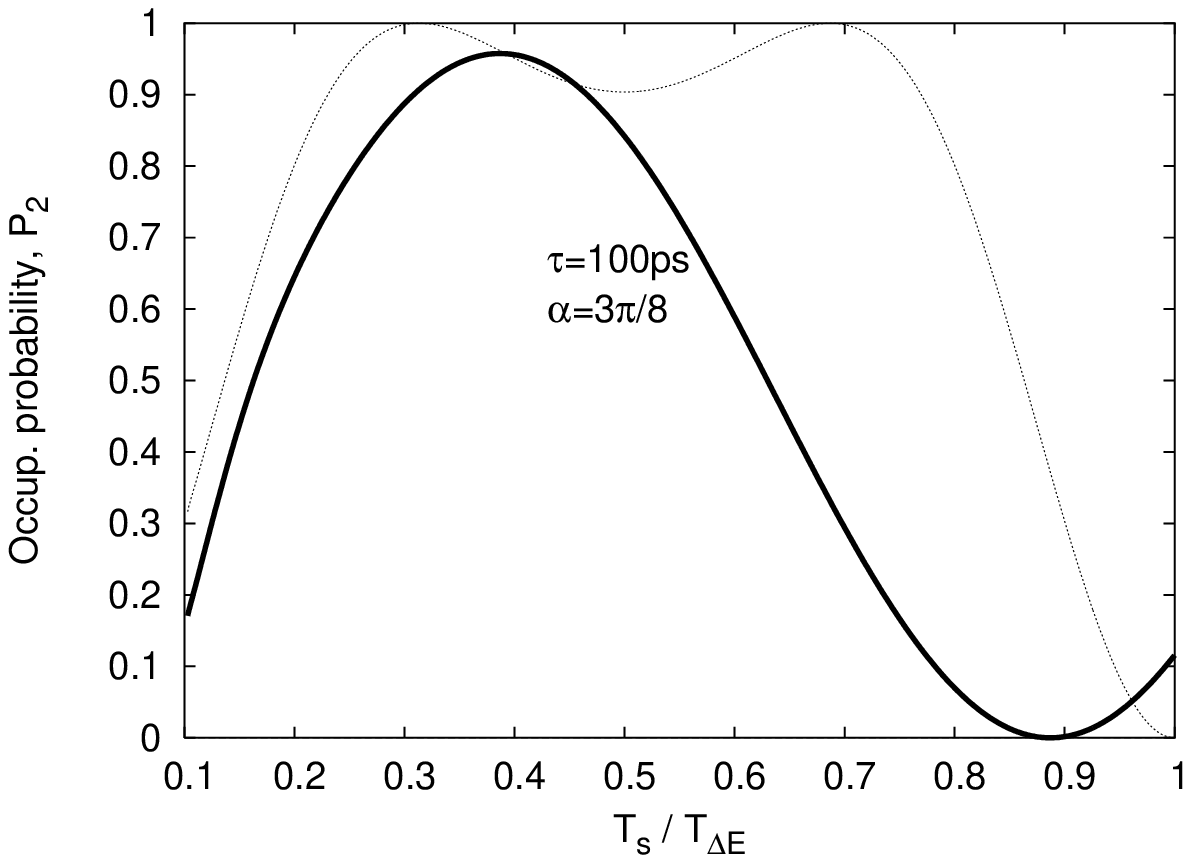}}}
\caption{Target state probability as a function of the separation time $T_s =
T_2 - T_1$ between the two pulses, for integrated pulse strength $\alpha =
\pi/2$, $\pi/4$, and $3 \pi/8$.  On the left the pulse width is a rather narrow
10 ps, while on the right the pulse width is 100 ps, where the kicked
approximation is breaking down.  Again the heavy line denotes the exact result
including time ordering, while the thin dashed line denotes the intermediate
picture result without time ordering.  The Schr\"odinger picture result without
time ordering is identically zero for all values of $T_s$ and $\alpha$.}
\end{figure}

Finally, we consider the role of time ordering in the case of two equal and
opposite pulses separated by an interval $T_s$.  In the limit of ideal kicks,
the kick-antikick evolution operator has been expressed analytically above with
and without time ordering in Eqs.~(\ref{kickakick}) and (\ref{U0Idouble}).
This yields for the probability $P_2(T_f)$ at times $T_f$ after the second
kick,
\begin{eqnarray}
\label{Pdoublecomp}
  P_2(T_f) &=& \sin^2{\gamma T_s} \sin^2 {2 \alpha}+O(\beta^2) \nonumber \\
  P_{I \, 2}^0(T_f) &=& \sin^2 \left [2 \alpha e^{-\beta^2}
   \sin^2 (\gamma T_s) \right] \\
  P_2^0(T_f) &=& 0 \nonumber  \ \ .
\end{eqnarray}  
The existence of these analytic results is helpful in studying the role of time
ordering in $P_2$.

In Fig.~5 we compare $P_2$ with $P_{I \, 2}^0$ as a function of the separation
time $T_s$ between the pulses.  Here the measurement time $T_f$ is taken to be
well after the second pulse has decayed, $T_f - T_2 \gg \tau$. We note that, as
expected, in all cases the occupation probability goes to zero as the two
opposite pulses coalesce, i.e. as $T_s \to 0$.  On the left side we show the
occupation probability for a Gaussian pulse of width $\tau$ = 10 ps for three
values of the pulse strength $\alpha$, namely $\alpha = \pi/2$ corresponding to
a kick that turns a qubit from off to on at $T_1$ and back to off at $T_2$,
$\alpha = \pi/4$ where a qubit is on after $T_2$, and an intermediate strength
$\alpha = 3 \pi/8$.  On the left side $\tau/T_{\Delta E}= 10^{-2}$, so that the
kicked result is accurately obtained.  We note that $P_{I \, 2}^0  \approx
P_2$, and time ordering effects vanish, when $T_s /T_{\Delta E}= \gamma T_s/pi$
is an integer or half-integer, consistent with Eq.~(\ref{Pdoublecomp}).  Away
from these special values, large time ordering effects are present.  On the
right side, the pulses are quite broad ($\tau/T_{\Delta E}= 10^{-1}$), and the
kicked approximation is clearly breaking down.  

\section{Discussion}

In this paper we have studied the role of time ordering in a strongly perturbed
two-state quantum system.  We have defined the time ordering effect as the
difference between a calculation with time ordering and one without time
ordering.  This is the way correlation, entanglement, and non-random processes
are also defined~\cite{mw}.  In all of these cases it is useful to define
carefully the limit without the effect (time ordering, correlation, or
non-randomness).  In the case of time ordering we have seen that the limit
without time correlation depends on how one separates $\hat H$ into $\hat H_0 +
\hat V$, i.e. what we call a choice of gauge.  In practice this choice
sometimes rests on the choice of the time averaged interaction.  Precisely this
same issue occurs in defining correlation~\cite{mcbook}, namely the somewhat
arbitrary choice of a mean field interaction.  A similar problem can arise in
defining entanglement~\cite{mabuchi} and non-random processes.  In any case the
problem is not new.

For clarity and simplicity we have chosen $\hat H_0 = {\rm const} \times
\sigma_z$ and $\hat V = f(t) V_0 \sigma_x$ in this paper.  However other gauge
choices are possible and in some cases may be more sensible.  Specifically, in
many cases experimental conditions can lead to a sensible gauge choice where,
for example, the asymptotic state of an unperturbed atom is an eigenstate of
$\hat H_0$, and $\hat V$ corresponds to an external electric or magnetic field
imposed on the atom during part of the experiment.  Such a $\hat V$ may
sensibly contain $\sigma_z$ and/or $\sigma_y$ components.

We have shown above that the limit of no time ordering depends on the picture
(representation) used.  While this difference is small when time ordering
effects are small, the differences can be large otherwise.  Hence it might be
argued that our time ordering analysis is useful primarily  to determine if
time ordering effects are small or large.  This argument sometimes occurs in
the use of correlation (although it is not common to use different
representations for correlation).  In any case the dependence on representation
is not new~\cite{converg}.  Convergence properties of the Magnus expansion in
the Schr\"odinger and interaction pictures have been known for some time to
differ widely~\cite{salzman}.  While the Schr\"{o}dinger picture result (that
in a sense corresponds to an especially simple gauge choice of $\hat H_0 = 0$)
is formally easier to write down, it often does not yield predictions as
reliable as the more complete results found in the intermediate picture.  We
have illustrated this above with calculations and analysis of $\hat U^0$ and
$\hat U^0_I$ for single and multiple pulses.  The intermediate picture takes
maximum advantage of knowledge about the eigenstates and spectrum of $\hat
H_0$.  In other words the intermediate picture is generally more complete than
the Schr\"{o}dinger picture and often more sensible.  It is useful to separate
$\hat H$ into $\hat H_0 + \hat V$ in such a way as to include as much of the
problem as possible in $\hat H_0$, whose solutions are known.  In the extreme
limit of the Heisenberg picture (which can be thought of as a gauge choice
where $\hat H = \hat H_0$ and $\hat V =0$), we have $\hat U_H = \hat U_H^0 =
1$, and there is never any time ordering in the time evolution.

Here we have worked in the time domain and formulated the question of time
ordering by explicitly working with the Dyson time ordering operator, $T$.
Equivalently one may work in the energy domain, as has been done recently in
the context of atomic scattering to analyze experimental data and identify time
ordering effects~\cite{gm03,bruch}.  A key transformation for time ordering
from the time domain to the energy domain is the Fourier transform of the step
function, namely, $\int e^{i E' t} \Theta(t) e^{-i E t} dt = \frac{1}{E - E' +
i\eta} = \pi \delta(E - E') + i P_v \frac{1}{E-E'}$, where the effect of time
ordering is associated with $i \eta$, which gives rise to the principal value
term.  The $i \eta$ carries the effect of the boundary condition on the Green's
function in energy space.  This is discussed in more detail
elsewhere~\cite{gm01,gm03}.  Since $[\hat H(t''),\hat H(t')]$ provides a
connection between interactions at $t''$ and $t'$, the quantum time propagator
includes non-local effects in time.  An example of a counter-intuitive time
sequence occurs in stimulated Raman adiabatic passage (STIRAP), where efficient
and robust population transfer is attained using two pulsed radiation fields in
a three-level system~\cite{bergmann}.

The time ordering effects considered in this paper are associated with a
sequential ordering of interactions.  The normal boundary condition imposed on
the evolution operator is that the sequence proceeds in the direction of
increasing time (or alternatively decreasing time to study time reversal).
Hence effects of time ordering are associated with a direction of the flow of
time.  In this paper we have not included dissipation; hence all amplitudes
explicitly satisfy invariance under time reversal, e.g. in
Eqs.~(\ref{kickedut}) and (\ref{kickakick}) which include effects due to time
ordering.  Effects of time ordering have been observed in systems that satisfy
time reversal invariance~\cite{zhao,aarhus,bruch}.  This means that observable
evidence of the direction of time (time ordering) can be obtained without
violating the symmetry of time reversal invariance.

\section{Summary}  

In this paper we have given a definition of time ordering in a strongly
perturbed quantum system, namely that time ordering is the difference between
calculations with and without time ordering.  This definition is similar to the
definition of correlation.  In both cases effects arise from differences
between an instantaneous interaction and its averaged value.  When the effect
of time ordering is small, the dependence on representation is weak.  However,
when time ordering effects are large, the difference between representations
can also be large.  We have considered in detail time ordering for qubits that
are strongly and suddenly perturbed by an external interaction.  We have
illustrated our methods for a $2s - 2p$ transition in atomic hydrogen caused by
a Gaussian pulse of finite width in time.  Other diabatically changing qubits
may also be analyzed with our methods.  Simple analytic expressions have been
given for the occupation amplitudes and probabilities for kicked qubits,
including single and multiple kicks.  We think that it should be possible to
find analytic solutions for correlated kicked qubits, so that time coupled
interacting qubits may also be studied analytically.

\begin{acknowledgments}
We thank D. Uskov, C. Rangan, B. Shore, J. Eberly, and P. Berman for
useful discussions.  A. B. is supported by TAMS GL Fund No. 211093 through
Tulane University.
\end{acknowledgments}


\begin{thebibliography}{99}
\bibitem{rwa} The rotating wave approximation (RWA) has been extensively and
sensibly used.  However, this approach is not well suited for external
kicks that require a broad frequency spectrum.
\end{thebibliography}
\end{document}